\documentclass[aps,prb,twocolumn,groupedaddress,amsmath,amssymb,showpacs]{revtex4}

\usepackage{amssymb}
\usepackage{amsmath}
\usepackage{graphicx}
\usepackage{subfigure}
\usepackage{textcomp}
\usepackage{color}
\usepackage{amsfonts}
\usepackage{bbold}
\usepackage{dsfont}
\usepackage{epsfig}

\newcommand{\arctanh}[1]{\operatorname{arctan}}

\begin{document}

%\preprint{}


\title{Accurate self-energy algorithm for quasi-1D systems}



\author{Ivan Rungger}
\author{Stefano Sanvito}

%\thanks{}

\affiliation{School of Physics and CRANN, Trinity College, Dublin 2, Ireland}


\date{December 9, 2007}

\begin{abstract}
We present a complete prescription for the numerical calculation of surface Green's functions and self-energies of semi-infinite quasi-onedimensional systems. Our work extends the results of Sanvito \textit{et al.} [\onlinecite{stefanosene}] generating a robust algorithm to be used in conjunction with {\it ab initio} electronic structure methods. We perform a detailed error analysis of the scheme and find that the highest accuracy is found if no inversion of the usually ill conditioned hopping matrix is involved. Even in this case however a transformation of the hopping matrix that decreases its condition number is needed in order to limit the size of the imaginary part of the wave-vectors. This is done in two different ways, either by applying a singular value decomposition and setting a lowest bound for the smallest singular value, or by adding a random matrix of small amplitude. By using the first scheme the size of the Hamiltonian matrix is reduced, making the computation considerably faster for large systems. For most energies the method gives high accuracy, however in the presence of surface states the error diverges due to the singularity in the self-energy. A surface state is found at a particular energy if the set of solution eigenvectors of the infinite system is linearly dependent. This is then used as a criterion to detect surface states, and the error is limited by adding a small imaginary part to the energy. 
\end{abstract}

%\keywords{}

\maketitle

\section{\label{sec:intro}Introduction}

The electronic transport properties of quasi-onedimensional (1D) systems, described by a localized orbitals basis set, can be calculated using the nonequilibrium Green's function (NEGF) method.\cite{datta,smeagol1,transiesta,guotaylor} The system is usually divided into two semi-infinite left- and right-hand side leads, and a scattering region joining them. The effect of the leads onto the scattering region is taken into account by the so called self-energies (SE), which can be calculated from the surface Green's function (SGF) of the semi-infinite leads. These can be obtained either with recursive methods \cite{kudrnovsky1,kudrnovsky2,lopez,nardelli} or by using a semi-analytic formula.\cite{stefanosene,guotaylor,umerski,ando1,butlersene} Recursive methods are affected by poor convergence for some critical systems, typically when the Hamiltonian for the leads is rather sparse. Semi-analytical methods instead bypass those problems by construction, however major difficulties arise if the hopping matrices are singular or, more generally, ill conditioned. Unfortunately the condition of the Hamiltonian is set by the electronic structure of the leads and by the unit cell used, and thus it is largely not controllable. For this reason an algorithm that performs under the most generic conditions is highly desirable. Here we present an improved semi-analytical method that overcomes these limitations and thus represents a robust algorithm for quantum transport based on {\it ab initio} electronic structure. 

In the first part of the paper the extended algorithm for the calculation of the SE is presented. First the construction of the Green's function of an infinite 1D system as derived in reference [\onlinecite{stefanosene}] is recast into a more general form based on the notion of a complex group velocity. Then we present an extension of such method to the calculation of the SGF and SE. The new algorithm is defined also for the case of singular hopping matrices. This largely improves the numerical accuracy. However we find that even such an improved scheme sometimes fails if the hopping matrices are close to being singular. We overcome this problem by performing a transformation of the hopping matrix that reduces its condition number $\kappa$, defined as the ratio between its largest to its smallest singular value.\cite{ttao,higham} This transformation limits the maximum absolute value of the imaginary part of the Bloch wave-vectors, increasing both accuracy and stability. Two approaches are presented, the first is based on a singular value decomposition (SVD), which is also used to significantly reduce the dimension of the Hamiltonian, while the second consists in adding a random noise matrix. This extended scheme is implemented in the NEGF \textit{ab initio} transport code \textit{Smeagol},\cite{smeagol1,smeagol2} based on the density functional theory (DFT) code SIESTA.\cite{siesta}

In the second part of this work we present  three examples of calculations performed with our new implementation. We compare the results to the ones obtained by using the original method of reference [\onlinecite{stefanosene}], finding a considerable improvement. However, although the algorithm appears very robust, our detailed error analysis reveals that for a given system the accuracy is lost at some specific energies. This is caused by the divergence of one of the SE eigenvalues. The physical origin of this behavior lies in the presence of surface states very weakly coupled to the semi-infinite leads. Surface states appear whenever at a given energy the set of Bloch functions (with both real and imaginary wave-vectors) for the infinite quasi-1D system is linearly dependent. In the simplest case this corresponds to two Bloch functions being equal inside the unit cell. A small imaginary part is thus added to the energy in a small energy range around the surface state. It is shown that this has little effect on the transport properties in the high transmission regime, whereas for low transmission it has a substantial influence on the results. Crucially only a very small imaginary part is used, and moreover this is added only around the energy of the surface state and thus the error can be carefully controlled.

\section{Retarded Green's function for an infinite system}
\begin{figure}
\center
\includegraphics[width=7.0cm,clip=true]{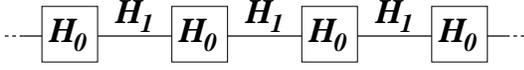}
\caption{Schematic representation of the system with onsite Hamiltonian $H_0$ and hopping $H_1$. The overlap matrix has the same structure.}
\label{fig:onedimh}
\end{figure}

Following the scheme introduced in reference [\onlinecite{stefanosene}] the construction of the retarded Green's function for an infinite quasi-1D system is now recalled. This is the starting point for the calculation of the SGF. It is assumed that the Hamiltonian is written over a localized orbitals basis set and that the interaction has finite range. The size of the unit cell can be chosen to guarantee interaction only to the first nearest neighboring unit cells. The total Hamiltonian of the system $H_{zz'}$ (the integers $z$ and $z'$ label the unit cells) can then be written as
\begin{equation}
H_{zz'}=H_0~\delta_{zz'} + H_1~\delta_{z,z'-1} + H_{-1}~\delta_{z,z'+1},
\label{eq:ham}
\end{equation}
where $H_0, H_1$ and $H_{-1}$ are $N\times N$ matrices, with $N$ being the number of orbitals comprised in the unit cell (see figure \ref{fig:onedimh}).
%*****************************************************************
%*****************************************************************
If time-reversal symmetry holds then $H_{0}=H_0^{\dagger}$, and $H_{-1}=H_1^{\dagger}$, however the solutions presented here are valid also in the more general case when $H_{0} \ne H_0^{\dagger}$ and/or $H_{-1}\ne H_1^{\dagger}$. We further assume that the overlap matrix $S_{zz'}$ has the same structure and range of the Hamiltonian
\begin{equation}
S_{zz'}=S_0~\delta_{zz'} + S_1~\delta_{z,z'-1} + S_{-1}~\delta_{z,z'+1},
\label{eq:overlap}
\end{equation}
where $S_0, S_1$ and $S_{-1}$ are again $N \times N$ matrices with the same meaning of their Hamiltonian counterparts.

\subsection{Bloch states expansion}
The solutions of the Hamiltonian equation for the associated infinite periodic system $\sum_{z'} H_{zz'}~\psi_{z'}=E~\sum_{z'} S_{zz'}~\psi_{z'}$ are Bloch functions $\psi_z=e^{i k z} \phi$, where $\psi_z$ and $\phi$ are $N$-dimensional vectors and $k$ is the wave-vector, which in general is a complex number. For a given real or complex energy $E$ there are $2 N$ solutions with wave-vectors $k_n$ and corresponding wavefunctions $\phi_n$. Each of them satisfies
\begin{eqnarray}
\left(H_0 + H_1 e^{i k_n} + H_{-1} e^{-i k_n}\right)\phi_n&=&\nonumber\\
E \left(S_0 + S_1 e^{i k_n} +\right.& S_{-1}&\left. e^{-i k_n}\right) \phi_n.
\end{eqnarray}
If we define $K_\alpha=H_\alpha-ES_\alpha$, ($\alpha=-1,0,1$), the equation above can be rewritten as
\begin{equation}
\left(K_0 + K_1 e^{i k_n} + K_{-1} e^{-i k_n}\right)\phi_{\mathrm{R},n}=0,
\label{eq:scheq}
\end{equation}
where the additional index R denotes explicitly that the solution is a right eigenvector. The corresponding left eigenvector $\phi_{\mathrm{L},n}$ satisfies
\begin{equation}
\phi_{\mathrm{L},n}^{\dagger}\left(K_0 + K_1 e^{i k_n} + K_{-1} e^{-i k_n}\right)=0.
\label{eq:scheql}
\end{equation}
Time-reversal symmetry gives $\phi_{\mathrm{L},n}=\phi_{\mathrm{L}}(k_n)=\phi_\mathrm{R}(k_n^*)$, so that in the case of real $k_n$ (propagating states) left and right eigenvectors are equal. For complex $k_n$ left and right eigenvectors are different, describing left- and right-decaying states. The sets $\lbrace k_n\rbrace$, $\lbrace \phi_{\mathrm{R},n}\rbrace$ and $\lbrace\phi_{\mathrm{L},n}\rbrace$ that satisfy eqs. (\ref{eq:scheq}) and (\ref{eq:scheql}) at a given energy can be found by solving a quadratic eigenvalue problem\cite{tisseur,chunhuaguo} of the form
\begin{eqnarray}
\left( \begin{array}{cc}
-K_0 &-K_{-1} \\
\mathbb{1}   &\mathbb{0}
\end{array} \right) \Phi_{\mathrm{R},n}&=&
e^{i k_n}
\left( \begin{array}{cc}
K_1 &\mathbb{0} \\
\mathbb{0}   &\mathbb{1}
\end{array} \right)\Phi_{\mathrm{R},n},
\label{eq:qep1}\\
\Phi_{\mathrm{L},n}^\dagger
\left( \begin{array}{cc}
-K_0 &-K_{-1} \\
\mathbb{1}   &\mathbb{0}
\end{array} \right) &=&
e^{i k_n}
\Phi_{\mathrm{L},n}^\dagger
\left( \begin{array}{cc}
K_1 &\mathbb{0} \\
\mathbb{0}   &\mathbb{1}
\end{array} \right),
\label{eq:qep2}
\end{eqnarray}
where
\begin{eqnarray}
&\Phi_\mathrm{R,n}=&
\left( \begin{array}{c} e^{i \frac{k_n}{2}} \\ e^{-i \frac{k_n}{2}} \end{array} \right)\frac{\phi_{\mathrm{R},n}}{\sqrt{v_n}},\\
&\Phi_\mathrm{L,n}^\dagger=&
\frac{i\phi_{\mathrm{L},n}^\dagger}{\sqrt{v_{n}}}
\left( \begin{array}{cc} e^{i \frac{k_n}{2}} &, - e^{-i \frac{k_n}{2}} K_{-1} \end{array} \right).
\end{eqnarray}
Here $\mathbb{1}$ and $\mathbb{0}$ are respectively the $N\times N$ unit and zero matrices. The normalization constant is the square root of the complex group velocity $v_n=\partial E/\partial k_n$ ($\hbar =1$) equal to
\begin{eqnarray}
v_n&=&\frac{i}{l_n}\phi_{\mathrm{L},n}^\dagger\left( K_1 e^{ik_n} -e^{-ik_n}K_{-1} \right) \phi_{\mathrm{R},n},\\
l_n&=&\phi_{\mathrm{L},n}^\dagger\left( S_0+S_1 e^{ik_n} +S_{-1}e^{-ik_n} \right) \phi_{\mathrm{R},n}.
\label{eq:ln}
\end{eqnarray}
In the following we assume that the eigenvectors $\phi_{\mathrm{R},n}$ and $\phi_{\mathrm{L},n}$ are always normalized to give $l_n=1$. If time-reversal symmetry holds then $v(k_n^*)=v_{n}^*$, so that if the imaginary part of $k_n$ is zero the group velocity is real. Note that, at variance with reference [\onlinecite{stefanosene}], eqs. (\ref{eq:qep1}) and (\ref{eq:qep2}) avoid the inversion of $K_1$, so that they eliminate a possible source of singularities in the calculation of $k_n$, $\phi_{\mathrm{R},n}$ and $\phi_{\mathrm{L},n}$.

The full sets of left $\lbrace{\Phi_{\mathrm{L},n}}\rbrace$ and right eigenvectors $\lbrace{\Phi_{\mathrm{R},n}}\rbrace$ form a complete and orthogonal basis. The orthogonality relation is
\begin{equation}
\Phi_{\mathrm{L},n}^\dagger
\left( \begin{array}{cc} K_1 &\mathbb{0} \\ \mathbb{0} &\mathbb{1} \end{array} \right)
 \Phi_{\mathrm{R},m} = c_n \delta_{nm},
\label{eq:orthor}
\end{equation}
where $c_n$ is a constant. This leads to
\begin{equation}
i\phi_{\mathrm{L},n}^\dagger\left( K_1 e^{ik_n} -e^{-ik_m}K_{-1} \right) \phi_{\mathrm{R},m}=
v_n ~ c_n~\delta_{nm}.
\end{equation}
For $n=m$ this equation is only satisfied if $c_n=1$, in which case it corresponds to the definition of $v_n$. With the chosen normalization the basis is therefore orthonormal. The corresponding completeness relation then reads
\begin{equation}
\sum_{n=1}^{2N} \Phi_{\mathrm{R},n}\Phi_{\mathrm{L},n}^\dagger\left( \begin{array}{cc} K_1 &\mathbb{0} \\ \mathbb{0} &\mathbb{1} \end{array} \right)
=\left( \begin{array}{cc} \mathbb{1} &\mathbb{0} \\ \mathbb{0} &\mathbb{1} \end{array} \right),
\label{eq:ortho}
\end{equation}
and provides the three following useful relations
\begin{eqnarray}
\sum_{n=1}^{2N} \frac{\phi_{\mathrm{R},n} \phi_{\mathrm{L},n}^+}{v_n} = \mathbb{0}
\label{eq:rel1},\\
K_1\sum_{n=1}^{2N} i e^{i k_n} \frac{\phi_{\mathrm{R},n} \phi_{\mathrm{L},n}^+}{v_n} =\mathbb{1},
\label{eq:rel2}
\\
K_{-1}\sum_{n=1}^{2N} -ie^{-i k_n} \frac{\phi_{\mathrm{R},n}\phi_{\mathrm{L},n}^+}{v_n} =\mathbb{1}.
\label{eq:rel3}
\end{eqnarray}
Note that in eqs. (\ref{eq:ortho}-\ref{eq:rel3}) the sums run over all $2N$ solutions.
If $K_1=K_{-1}^\dagger$ and $K_0=K_0^\dagger$ eqs. (\ref{eq:rel2}) and (\ref{eq:rel3}) are equivalent.

\subsection{Green's function}

The retarded Green's function $g_{zz'}$ of the system is defined by means of the Green's equation
\begin{equation}
\sum_{z'} g_{zz'} \left[\left(E+i\delta\right) S_{z'z''} - H_{z'z''}\right]=\delta_{zz''},
\label{eq:gf}
\end{equation}
with $\delta \rightarrow 0^{+}$ real. In what follows we present and expand, by using left and right Bloch functions, the solution to eq. (\ref{eq:gf}) given in reference [\onlinecite{stefanosene}] only in terms of the right eigenvectors $\phi_\mathrm{R}$. First we divide the $2N$ $\phi_{\mathrm{R},n}$ vectors into $N$ right-going states with either Im$(k_n)>0$ (right decaying) or Im$(k_n)=0$ and $v_n>0$ (right propagating), and $N$ left-going states with either Im$(k_n)<0$ (left decaying) or Im$(k_n)=0$ and $v_n<0$ (left propagating). As a matter of notation in order to distinguish left- from right-going states, in what follows we indicate the right-going states with $k$, $\phi$ and $v$, and the left-going states with a bar over these quantities, i.e. $\bar{k}$, $\bar{\phi}$ and $\bar{v}$.

As in reference [\onlinecite{stefanosene}] we introduce the duals $\tilde{\phi}_{\mathrm{R},n}$ of the right-going states $\phi_{\mathrm{R},n}$ defined by $\tilde{\phi}_{\mathrm{R},n}^\dagger\phi_{\mathrm{R},m}=\delta_{nm}$, and the duals $\tilde{\bar{\phi}}_{\mathrm{R},n}$ of the left-going states $\bar{\phi}_{\mathrm{R},n}$ defined by $\tilde{\bar{\phi}}_{\mathrm{R},n}^\dagger\bar{\phi}_{\mathrm{R},m}=\delta_{nm}$. If we define the matrices $Q$ and $\bar{Q}$ as
\begin{equation}
\begin{split}
Q&=\left(\begin{array}{ccccc}\phi_{\mathrm{R},1}&\phi_{\mathrm{R},2}&\dots&\phi_{\mathrm{R},N}\end{array}\right),\\
\bar{Q}&=\left(\begin{array}{ccccc}\bar{\phi}_{\mathrm{R},1}&\bar{\phi}_{\mathrm{R},2}&\dots&\bar{\phi}_{\mathrm{R},N}\end{array}\right),
\label{eq:Q}
\end{split}
\end{equation}
then the duals can be obtained by simple inversion
\begin{equation}
\begin{split}
\left(\begin{array}{ccccc}\tilde{\phi}_{\mathrm{R},1}&\tilde{\phi}_{\mathrm{R},2}&\dots&\tilde{\phi}_{\mathrm{R},N}\end{array}\right)&=\left(Q^{-1}\right)^\dagger,\\
\left(\begin{array}{ccccc}\tilde{\bar{\phi}}_{\mathrm{R},1}&\tilde{\bar{\phi}}_{\mathrm{R},2}&\dots&\tilde{\bar{\phi}}_{\mathrm{R},N}\end{array} \right)&=\left(\bar{Q}^{-1}\right)^\dagger.
\label{eq:adjbasisl}
\end{split}
\end{equation}
The inversions in eqs. (\ref{eq:adjbasisl}) are usually well defined, unless $Q$ and $\bar{Q}$ do not have full rank.  We will return on this aspect in section \ref{sec:sstates}, for the moment we assume that the duals can always be constructed.

The Green's function calculated in reference [\onlinecite{stefanosene}] is then
\begin{equation}
g_{zz'}=\left\{ \begin{array}{c}
\sum_{n=1}^N \phi_{\mathrm{R},n} e^{i k_n (z-z')} \tilde{\phi}_{\mathrm{R},n}^\dagger V^{-1}~~z\ge z'\\
\,\sum_{n=1}^N \bar{\phi}_{\mathrm{R},n} e^{i \bar{k}_n (z-z')} \tilde{\bar{\phi}}_{\mathrm{R},n}^\dagger V^{-1}~~z\le z',
\end{array} \right.
\label{eq:gfstefano}
\end{equation}
with the matrix $V=g_{zz}^{-1}=g_{00}^{-1}$ given by
\begin{equation}
V=K_{-1}\left(\sum_{n=1}^{N}  e^{-i k_n} \phi_{\mathrm{R},n} \tilde{\phi}_{\mathrm{R},n}^\dagger-
\sum_{n=1}^{N}  e^{-i \bar{k}_n} \bar{\phi}_{\mathrm{R},n} \tilde{\bar{\phi}}_{\mathrm{R},n}^\dagger \right).
\label{eq:v}
\end{equation}
We now introduce the right transfer matrices\cite{lopez,lopez2,nardelli} $T_\mathrm{R}$ and $\bar{T}_\mathrm{R}$
\begin{eqnarray}
T_\mathrm{R}&=&\sum_{n=1}^N \phi_{\mathrm{R},n} e^{i k_n} \tilde{\phi}_{\mathrm{R},n}^\dagger,
\label{eq:a}\\
\bar{T}_\mathrm{R}&=&\sum_{n=1}^N \bar{\phi}_{\mathrm{R},n} e^{-i \bar{k}_n} \tilde{\bar{\phi}}_{\mathrm{R},n}^\dagger.
\label{eq:bara}
\end{eqnarray}
Note that both $T_\mathrm{R}$ and $\bar{T}_\mathrm{R}$ have eigenvalues with complex modulus $\le 1$. For an integer $z$ the following relations hold
\begin{equation}
\begin{split}
\left(T_\mathrm{R}\right)^{z}=&\sum_{n=1}^N \phi_{\mathrm{R},n} e^{i k_n z} \tilde{\phi}_{\mathrm{R},n}^\dagger, \\
\left(\bar{T}_\mathrm{R}\right)^{z}=&\sum_{n=1}^N \bar{\phi}_{\mathrm{R},n} e^{- i \bar{k}_n z} \tilde{\bar{\phi}}_{\mathrm{R},n}^\dagger,
\end{split}
\end{equation}
which allow us to write the Green's function of equation (\ref{eq:gfstefano}) as
\begin{equation}
g_{zz'}=\left\{ \begin{array}{c}
\left(T_\mathrm{R}\right)^{z-z'} g_{00}~~~~~~z\ge z' \\
\left(\bar{T}_\mathrm{R}\right)^{z'-z} g_{00}~~~~~~z\le z'
\end{array} \right..
\label{eq:gfA}
\end{equation}
In the same way $V$ is rewritten as
\begin{equation}
V=g_{00}^{-1}=K_{-1}\left(T_\mathrm{R}^{-1}-\bar{T}_\mathrm{R}\right).
\label{eq:g00a}
\end{equation}
Note that although the matrices $T_\mathrm{R}$ and $\bar{T}_\mathrm{R}$ are in general well defined, the inverse of these matrices is not. In fact if $K_1$ and $K_{-1}$ are singular there are some $k_n$ with Im$(k_n)\rightarrow \infty$, so that $e^{i k_n}=0$ (see section \ref{sec:svdreduce}). In this case $T_\mathrm{R}$ does not have full rank and is therefore singular. The same argument holds for $\bar{T}_\mathrm{R}$. Equation (\ref{eq:g00a}) can therefore be used only if the matrices $K_1$ and $K_{-1}$ are not singular.

A possible way for overcoming such limitation is by using an equivalent form for the Green's function based on the left and right eigenvectors. The starting point is the relation (\ref{eq:rel1}) that will allow us to find the connection between the duals and the left eigenvectors. Eq. (\ref{eq:rel1}) contains a sum over both left- and right-going states. By moving the contribution of the left-going states to the right side of the equation we obtain $\sum_{n=1}^{N} \frac{\phi_{\mathrm{R},n} \phi_{\mathrm{L},n}^+}{i v_n} = -\sum_{n=1}^{N} \frac{\bar{\phi}_{\mathrm{R},n} \bar{\phi}_{\mathrm{L},n}^+}{i \bar{v}_n}=B$, where we have introduced the auxiliary matrix $B$. By multiplying $B$ from the left with either $\tilde{\phi}_\mathrm{R}^\dagger$ or $\tilde{\bar{\phi}}_\mathrm{R}^\dagger$ we obtain respectively $\tilde{\phi}_{\mathrm{R},n}^\dagger=\frac{1}{i v_n}\phi_{\mathrm{L},n}^\dagger B^{-1}$ and $\tilde{\bar{\phi}}_{\mathrm{R},n}^\dagger=\frac{1}{-i \bar{v}_n}\bar{\phi}_{\mathrm{L},n}^\dagger B^{-1}$. The matrix $B$ is determined by inserting these relations into eq. (\ref{eq:v}) and by using eq. (\ref{eq:rel3}), the result is $B=g_{00}$. The relation between the dual basis and the left eigenvectors is therefore
\begin{equation}
\tilde{\phi}_{\mathrm{R},n}^\dagger=\frac{1}{i v_n}\phi_{\mathrm{L},n}^\dagger g_{00}^{-1}~~,~~
\tilde{\bar{\phi}}_{\mathrm{R},n}^\dagger=\frac{1}{-i \bar{v}_n}\bar{\phi}_{\mathrm{L},n}^\dagger g_{00}^{-1}.
\label{eq:phitildeofphil}
\end{equation}
This result allows us to rewrite the Green's function of eq. (\ref{eq:gfstefano}) in a shorter form
\begin{equation}
g_{zz'}=\left\{ \begin{array}{c}
\sum_{n=1}^N ~\frac{1}{iv_n}\phi_{\mathrm{R},n} e^{i k_n (z-z')} \phi_{\mathrm{L},n}^\dagger~~~z\ge z' \\
\sum_{n=1}^N \frac{1}{-i\bar{v}_n}\bar{\phi}_{\mathrm{R},n} e^{i \bar{k}_n (z-z')} \bar{\phi}_{\mathrm{L},n}^\dagger ~~~z\le z'.
\end{array} \right.~~~
\label{eq:gfansatz}
\end{equation}
This result represents a generalization to complex energies and to systems breaking time-reversal symmetry of the solution given in references [\onlinecite{allen,chungchang}] for Hermitian Hamiltonians, real energy and an orthogonal tight binding model. This derivation shows that the Green's function can be equivalently expressed by using the right eigenvectors and their duals (eq. (\ref{eq:gfstefano})), or both the right and left eigenvectors (eq. (\ref{eq:gfansatz})). It is thus possible to move from one representation to the other through eq. (\ref{eq:phitildeofphil}) that relates the duals to the left eigenvectors. One can then decide which representation to use, depending on the specific problem investigated. We note that eq. (\ref{eq:gfansatz}) has the benefit that $g_{00}$ can be calculated also in the case where the two matrices $K_1$ and $K_{-1}$ are singular. For those $k_n$ where Im$(k_n)\rightarrow \infty$  the group velocity becomes $v_n=i~\phi_{\mathrm{L},n}^\dagger K_0 \phi_{\mathrm{R},n}$ and is therefore well defined ($\bar{v}_n=- i~\bar{\phi}_{\mathrm{L},n}^\dagger K_0 \bar{\phi}_{\mathrm{R},n}$ for Im$(\bar{k}_n)\rightarrow -\infty$).

As a matter of completeness we show that a representation entirely based on the left Bloch functions and their duals $\tilde{\phi}_{\mathrm{L},n}$ and $\tilde{\bar{\phi}}_{\mathrm{L},n}$ is also possible. By multiplying eq. (\ref{eq:gfansatz}) respectively by $\tilde{\phi}_{\mathrm{L},n}$ and $\tilde{\bar{\phi}}_{\mathrm{L},n}$ from the right we obtain the two relations
\begin{equation}
\tilde{\phi}_{\mathrm{L},n}=\frac{1}{i v_n}g_{00}^{-1}\phi_{\mathrm{R},n}~~,~~
\tilde{\bar{\phi}}_{\mathrm{L},n}=\frac{1}{-i \bar{v}_n}g_{00}^{-1}\bar{\phi}_{\mathrm{R},n} .
\label{eq:phitildeofphir}
\end{equation}
Again the left transfer matrices $T_\mathrm{\mathrm{L}}$ and $\bar{T}_\mathrm{L}$ are defined as
\begin{eqnarray}
T_\mathrm{L}&=&\sum_{n=1}^N \tilde{\phi}_{\mathrm{L},n} e^{i k_n} \phi_{\mathrm{L},n}^\dagger
\label{eq:al}\\
\bar{T}_\mathrm{L}&=&\sum_{n=1}^N \tilde{\bar{\phi}}_{\mathrm{L},n} e^{-i \bar{k}_n} \bar{\phi}_{\mathrm{L},n}^\dagger,
\label{eq:baral}
\end{eqnarray}
and the Green's function of eq. (\ref{eq:gfansatz}) can be rewritten as
\begin{equation}
g_{zz'}=\left\{ \begin{array}{c}
g_{00}~\left(T_\mathrm{L}\right)^{z-z'}~~~~~~z\ge z' \\
g_{00}~\left(\bar{T}_\mathrm{L}\right)^{z'-z} ~~~~~~z\le z'
\end{array} \right..
\label{eq:gleft}
\end{equation}
The structure of eq. (\ref{eq:gleft}) is the same as that of eq. (\ref{eq:gfA}), with the difference that now $g_{00}$ is multiplied to the left of the transfer matrix. Finally we extend eq. (\ref{eq:g00a}) and present four equivalent relations for the inverse of $g_{00}$
\begin{eqnarray}
g_{00}^{-1}&=&K_{-1}\left(T_\mathrm{R}^{-1}-\bar{T}_\mathrm{R}\right)=K_{1}\left(\bar{T}_\mathrm{R}^{-1}-T_\mathrm{R}\right)\nonumber\\
&=&\left(T_\mathrm{L}^{-1}-\bar{T}_\mathrm{L}\right)K_{-1}=\left(\bar{T}_\mathrm{L}^{-1}-T_\mathrm{L}\right)K_{1}.
\label{eq:g00}
\end{eqnarray}
The second of these relations can be shown by multiplying eq. (\ref{eq:rel2}) by $g_{00}^{-1}$ from the right and then by using eq. (\ref{eq:phitildeofphil}). In the same way the third and fourth equations can be obtained by multiplying equations (\ref{eq:rel2}) and (\ref{eq:rel3}) by $g_{00}^{-1}$ from the left.

In the following we will use mostly the quantities expressed in terms of the right eigenvectors only, however the same conclusions can be derived using the left eigenvectors.

\subsection{Density of states}

As an example of the use of the Green's function in the form of eq. (\ref{eq:gfansatz}) we determine the spectral function $A$ and the density of states (DOS) of the infinite quasi-1D system for the special case where the Hamiltonian and the overlap matrices are Hermitian. The spectral function is defined as\cite{datta}
\begin{equation}
A_{zz'}=i\left[g-g^\dagger\right]_{zz'}=i\left[g_{zz'}-\left(g_{z'z}\right)^\dagger\right].
\label{eq:azzg}
\end{equation}
The DOS $\rho_z$ projected on the unit cell $z$ then is
\begin{equation}
\rho_z=\frac{1}{2\pi}\mathrm{Tr}\left[\sum_{z'}A_{zz'} S_{z'z}\right].
\end{equation}
By using eq. (\ref{eq:overlap}) this becomes
\begin{equation}
\rho_z=\frac{1}{2\pi}\mathrm{Tr}\left[A_{zz} S_{0}+A_{z,z-1} S_{1}+A_{z,z+1} S_{-1}\right].
\label{eq:dosg}
\end{equation}
In general the main contribution originates from the first term in the sum, which can be interpreted as the onsite DOS $\tilde{\rho}_{z}$
\begin{equation}
\tilde{\rho}_{z}=\frac{1}{2\pi}\mathrm{Tr}\left[A_{zz} S_{0}\right].
\label{eq:onsitedos}
\end{equation}
We now calculate $A$ and $\rho$ for the special case where $K_{-1}=K_1^\dagger$ and $K_0=K_0^\dagger$. In this case for Im$(k_n)=0$ we have $\phi_{\mathrm{L},n}=\phi_{\mathrm{R},n}$, whereas if Im$(k_n)\ne0$ then $\phi_{\mathrm{L},n}=\phi_{\mathrm{L}}(k_n)=\bar{\phi}_{\mathrm{R}}(k_n^*)$. In the same way for Im$(\bar{k}_n)=0$ we have $\bar{\phi}_{\mathrm{L},n}=\bar{\phi}_{\mathrm{R},n}$, whereas if Im$(\bar{k}_n)\ne0$ then $\bar{\phi}_{\mathrm{L},n}=\bar{\phi}_{\mathrm{L}}(\bar{k}_n)=\phi_{\mathrm{R}}(\bar{k}_n^*)$.  Therefore for each right decaying state with Im$(k_n)>0$ there is a left decaying state with $\bar{k}_{n}=k_n^*$  and $v(\bar{k}_n)^*=v(k_n)$. Using these relations when inserting the Green's function of eq. (\ref{eq:gfansatz}) in the definition of $A_{zz'}$, the contribution from all the decaying states cancels out. The only remaining contributions come from the propagating states, also denoted as open channels. For these $k_n^*=k_n$, $\bar{k}_n=-k_n$ and $v(\bar{k}_n)=-v(k_n)$. With these constraints, and by using eq. (\ref{eq:gfansatz}), the spectral function becomes
\begin{equation}
A_{zz'}=
\sum_{n}^{N_\mathrm{open}} \frac{e^{i k_n (z-z')}}{v_n}\phi_{\mathrm{R},n}  \phi_{\mathrm{R},n}^\dagger +
\frac{e^{-i k_n (z-z')}}{v_n}\bar{\phi}_{\mathrm{R},n}  \bar{\phi}_{\mathrm{R},n}^\dagger ,
\label{eq:azz}
\end{equation}
where $N_\mathrm{open}$ is the number of open channels (number of Bloch functions at a given energy with real positive $k$ vector). If there are no open channels $A_{zz'}=0$ and the Green's function is Hermitian. Finally, by using eqs. (\ref{eq:azz}) and (\ref{eq:dosg}), and the fact that the eigenvectors are normalized so to give $l_n=1$ (see eq. (\ref{eq:ln})), the DOS at the site $z=0$ is simply
\begin{equation}
\rho_0=\frac{1}{\pi}\sum_{n}^{N_\mathrm{open}}\frac{1}{v_n}.
\end{equation}
This is the well known result for the DOS of infinite periodic 1D systems.\cite{landauerbuttiker}

\section{Surface Green's function and self-energy}
\label{sec:sgf}

The retarded Green's function $g_\mathrm{S}$ for a quasi-periodic system, where the left and right sides are separated at the position $z=0$ (the left-hand side part extends from $z=-\infty$ to $z=-1$ and the right-hand side part from $z=1$ to $z=\infty$, with no coupling between the cells at $z=-1$ and $z=1$), can be constructed from the Green's function $g$ for the infinite chain as demonstrated in reference [\onlinecite{stefanosene}]
\begin{equation}
g_{S,zz'}=g_{zz'} - g_{z0}~g_{00}^{-1}~g_{0z'}.
\end{equation}
The left-hand side SGF is then defined as $G_\mathrm{L}=g_{\mathrm{S},-1,-1}$, and the right SGF as $G_\mathrm{R}=g_{\mathrm{S},11}$. The SGF can be obtained by using eq. (\ref{eq:gfA}) 
\begin{eqnarray}
G_\mathrm{L}&=&\left(\mathbb{1}-\bar{T}_\mathrm{R}T_\mathrm{R}\right)g_{00},\nonumber\\
G_\mathrm{R}&=&\left(\mathbb{1}-T_\mathrm{R}\bar{T}_\mathrm{R}\right)g_{00}.
\label{eq:sgfstefano}
\end{eqnarray}
This corresponds to the form derived in reference [\onlinecite{stefanosene}]. This result can be simplified by using the relations (\ref{eq:g00}) for $g_{00}$ to 
\begin{eqnarray}
G_\mathrm{L}&=&\bar{T}_\mathrm{R}~K_1^{-1},\nonumber\\
G_\mathrm{R}&=&T_\mathrm{R}~K_{-1}^{-1}.
\label{eq:gfs}
\end{eqnarray}
These equations unfortunately are only defined if $K_1$ and $K_{-1}$ are not singular. The same problem however does not affect the left and right SE, $\Sigma_\mathrm{L}=K_{-1}~G_\mathrm{L}~K_{1}$ and $\Sigma_\mathrm{R}=K_{1}~G_\mathrm{R}~K_{-1}$,\cite{smeagol1} since they simply are
\begin{eqnarray}
\Sigma_\mathrm{L}&=&K_{-1} \bar{T}_\mathrm{R},
\label{eq:senel}\\
\Sigma_\mathrm{R}&=&K_{1} T_\mathrm{R}.
\label{eq:sener}
\end{eqnarray}
%
%*****************************************************************
\begin{figure*}
\center
\includegraphics[width=15.5cm,clip=true]{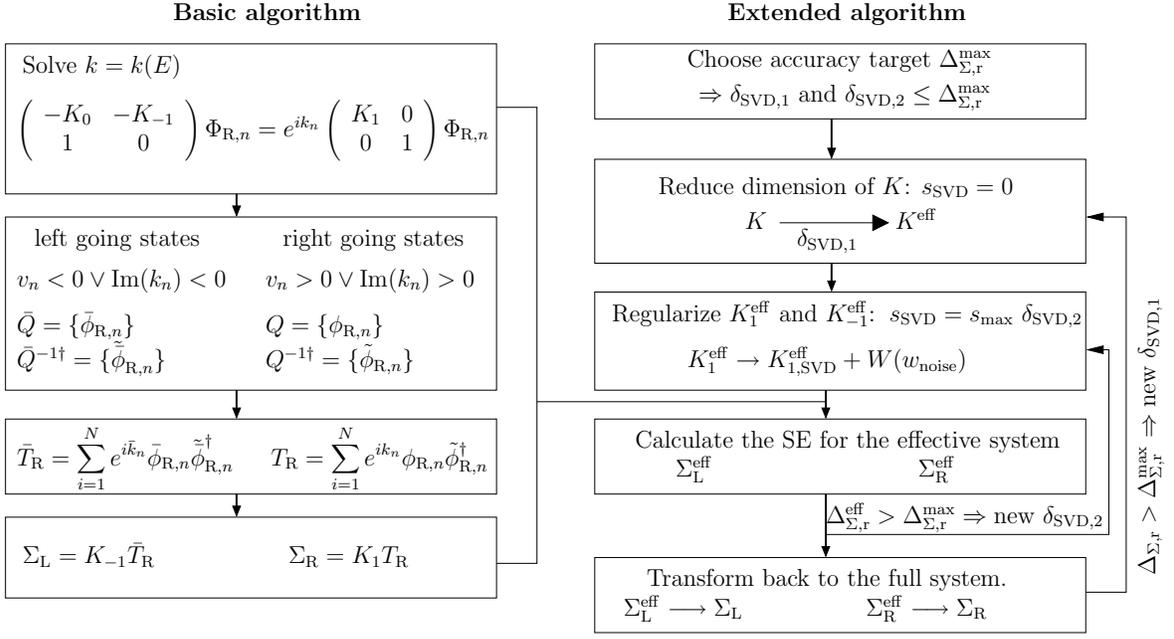}
\caption{Schematic diagram of the basic algorithm described in section \ref{sec:sgf} and of the extended algorithm described in section \ref{sec:ksvdnoise}.}
\label{fig:diagram}
\end{figure*}
%*****************************************************************
In complete analogy the same expressions obtained by using the left transfer matrices are $\Sigma_\mathrm{L}=T_\mathrm{L} K_{1}$ and $\Sigma_\mathrm{R}=\bar{T}_\mathrm{L} K_{-1}$. This result is equivalent to those obtained in references [\onlinecite{lopez,nardelli,ando1,umerski,butlersene}] and derived with different approaches, demonstrating the equivalence of those to our semi-analytical formula. Since NEGF-based transport codes simply require $\Sigma_\mathrm{L}$ and $\Sigma_\mathrm{R}$, our scheme allows the calculations of system with arbitrarily complicated electronic structure. A schematic tree diagram describing the steps involved in obtaining the SE is shown in figure \ref{fig:diagram} (basic algorithm).

Eqs. (\ref{eq:senel}) and (\ref{eq:sener}) demonstrate that the SE can be calculated directly without explicitly calculating $G_\mathrm{L}$ and $G_\mathrm{R}$. In situations where also the SGF are needed, these can be obtained by using the relation
\begin{eqnarray}
G_\mathrm{L} &=& - \left[K_0 +\Sigma_\mathrm{L} \right]^{-1},
\label{eq:rgfl}\\
G_\mathrm{R} &=& - \left[K_0 +\Sigma_\mathrm{R} \right]^{-1}.
\label{eq:rgfr}
\end{eqnarray}
This can be derived by adding one layer to the left and one to the right surfaces respectively.\cite{kudrnovsky1} In Appendix A we show that the SE calculated with eqs. (\ref{eq:senel}) and (\ref{eq:sener}) indeed fulfill the above equation. Moreover with the use of eqs. (\ref{eq:gfs}) and (\ref{eq:rgfl}) we can now regularize equation (\ref{eq:g00a}) also for the case where $T_\mathrm{R}$ is singular by writing it as
\begin{equation}
g_{00}^{-1}=-K_0-\Sigma_\mathrm{L}-\Sigma_\mathrm{R}.
\end{equation}
We have therefore a scheme where the SE are identified as the principal quantities, whereas the SGF and $g_{00}$ are derived from these.

When we compare the method of reference [\onlinecite{stefanosene}] with the equations derived above, we notice that now it is not necessary to calculate the matrix $g_{00}$ and its inverse using eq. (\ref{eq:g00a}) in order to obtain the SE. This is not defined in the case of singular $K_1$ and $K_{-1}$, and therefore we expect the new method to be more stable and accurate. Also the problems caused close to band edges by the Van Hove singularities in $g_{00}$ are avoided. Moreover the method in reference [\onlinecite{stefanosene}] relies on the calculation of the SGF in order to obtain the SE, whereas here the SGF is not needed. As we will show in section \ref{sec:sstates} close to surface states the error in the SGF is much larger than the one for the SE, so that we also expect a large improvement in the accuracy for those particular states. 

\section{Reducing the condition number of $K_1$ and $K_{-1}$}
\label{sec:ksvdnoise}

The accuracy with which the SE are calculated depends on the accuracy involved in solving eq. (\ref{eq:qep1}), a quadratic eigenvalue problem extensively studied in the past.\cite{tisseur,chunhuaguo} However most solution methods have problems if $K_1$ or $K_{-1}$ are close to being singular, or more generally if their condition number $\kappa$ is large. In this case some of the complex eigenvalues tend to infinity and others to zero at the same time, and this results in a loss of accuracy in numerical computations. When calculating $T_\mathrm{R}$ ($\bar{T}_\mathrm{R}$) however the contributions from the states with Im$(k_n) \rightarrow \infty$ (Im$(\bar{k}_n) \rightarrow -\infty$) are vanishingly small. It is therefore useful to limit the range of the eigenvalues $\lbrace e^{i k_n}\rbrace$ in such a way that the important eigenstates with small $|\mathrm{Im}(k_n)|$ and $|\mathrm{Im}(\bar{k}_n)|$ can be calculated accurately, while losing precision for the less important eigenstates with large $|\mathrm{Im}(k_n)|$ and $|\mathrm{Im}(\bar{k}_n)|$. In this section we show how this can be achieved by decreasing $\kappa(K_1)$ and $\kappa(K_{-1})$. Here we assume that $K_1=K_{-1}^\dagger$, so that $\kappa(K_1)=\kappa(K_{-1})$. Minor modifications are needed for the general case (see Appendix B).

In order to obtain $\kappa(K_1)$ first a SVD of the matrix is performed
\begin{equation}
K_1=U S V^\dagger.
\label{eq:svd}
\end{equation}
$U$ and $V$ are unitary matrices, and $S$ is a diagonal matrix, whose diagonal elements $s_n$ are the singular values. These are real and positive, and ordered so that $s_{n+1}\le s_n$. If $s_{\mathrm{max}}$ is the largest singular value, and $s_{\mathrm{min}}$  the smallest one, then the condition number is defined as $\kappa(K_1)=s_{\mathrm{max}}/s_{\mathrm{min}}$, with $K_1$ singular if $s_{\mathrm{min}}$ is zero.

We now replace $S$ with an approximate $S_{\mathrm{SVD}}$, whose diagonal elements $s_{\mathrm{SVD},n}$ are
\begin{equation}
s_{\mathrm{SVD},n}=\left\{ \begin{array}{c}
s_n             ~~~~~ s_n \ge s_{\mathrm{max}}~\delta_\mathrm{SVD} \\
s_{\mathrm{SVD}} ~~~  s_n <   s_{\mathrm{max}}~\delta_\mathrm{SVD}
\end{array} \right.,
\label{eq:ssvd}
\end{equation}
and accordingly $K_1$ with $K_{1,\mathrm{SVD}}=U S_{\mathrm{SVD}} V^\dagger$. The tolerance parameter $\delta_\mathrm{SVD}$ is a real positive number that determines the condition number of $K_{1,\mathrm{SVD}}$.

We now present two possible choices for $s_{\mathrm{SVD}}$. The first is to set $s_{\mathrm{SVD}}=0$, resulting in $K_{1,\mathrm{SVD}}$ being singular. We can then perform a unitary transformation in order to eliminate the degrees of freedom associated to $s_{\mathrm{SVD},n}=0$, and obtain an effective $K_1$ matrix ($K_1^{\mathrm{eff}}$) with reduced size for which $\kappa(K_1^{\mathrm{eff}}) \le \delta_\mathrm{SVD}^{-1}$. The second possibility is to set $s_{\mathrm{SVD}}=s_{\mathrm{max}}~\delta_\mathrm{SVD}$, so that by definition we have $\kappa(K_1) = \delta_\mathrm{SVD}^{-1}$. The accuracy obtained with both strategies is similar, the advantage of using the first however is that the size of the matrices is reduced, so that for big systems the computation is much faster. In our implementation we use both methods together, first we reduce the size of the system by setting $s_{\mathrm{SVD}}=0$, and then, if necessary, we further reduce the condition number for the effective system by limiting the smallest singular value.

\subsection{Reduction of system size}
\label{sec:svdreduce}
Here we set all the $M$ singular values $s_n$ smaller than $s_{\mathrm{max}}\delta_{\mathrm{SVD}}$ to zero, so that there are $N_{\mathrm{eff}}=N-M$ singular values $s_n$ with $s_n \ge s_{\mathrm{max}}\delta_{\mathrm{SVD}}$. The transformations needed in order to obtain the right SE are now presented (the procedure for the left SE is analogous). We apply the unitary transformation $K_{zz'}'=U^\dagger K_{zz'} U$, $\phi_{\mathrm{R},n}'=U^\dagger \phi_{\mathrm{R},n}$, and we define $K_1'=U^\dagger K_{1,\mathrm{SVD}} U$, $K_{-1}'=U^\dagger K_{-1,\mathrm{SVD}} U$, $K_{0}'=U^\dagger K_{0} U$. Since $M$ singular values of $K_{1,\mathrm{SVD}}$ are zero the transformed matrices have the structure
\begin{equation}
\begin{split}
K_1'=&
\left( \begin{array}{cc} K_{1,\mathrm{c}} &  K_{1,\mathrm{u}}  \\ \mathbb{0} & \mathbb{0}\end{array} \right),~~~~
K_{-1}'=
\left( \begin{array}{cc} K_{-1,\mathrm{c}} & \mathbb{0} \\  K_{-1,\mathrm{u}} & \mathbb{0}\end{array} \right), \\
K_{0}'=&
\left( \begin{array}{cc} A & B \\ C & D  \end{array} \right),~~~~~~~~~~~
\phi_{\mathrm{R},n}'=
\left( \begin{array}{c} \phi_{\mathrm{c},n}  \\ \phi_{\mathrm{u},n} \end{array} \right),
\end{split}
\label{eq:transfmat}
\end{equation}
where the dimensions of the new matrices are: $N_\mathrm{eff}\times N_\mathrm{eff}$ for $K_{1,\mathrm{c}}$, $K_{-1,\mathrm{c}}$ and $A$, $N_\mathrm{eff}\times M$ for $K_{1,\mathrm{u}}$ and $B$, $M\times N_\mathrm{eff}$ for $K_{-1,\mathrm{u}}$ and $C$, and $M\times M$ for $D$. Finally $\phi_{\mathrm{c},n}$ is a column vector of dimension $N_\mathrm{eff}$, and $\phi_{\mathrm{u},n}$ is of dimension $M$. The transformed form of eq. (\ref{eq:scheq}) is
\begin{equation}
\left(K_0' + K_1' e^{i k_n} + K_{-1}' e^{-i k_n}\right)\phi_{\mathrm{R},n}'=0.
\label{eq:scheqv}
\end{equation}
Due to the structure of $K_{-1}'$ there are $M$ solutions to this equation with $e^{i k_n}=0$ and $\phi_{\mathrm{c},n}=0$. We therefore split up the right-going states into those with finite $e^{i k_n}\ne 0$ and those with $e^{i k_n}=0$. For the first set, from eq. (\ref{eq:scheqv}), we obtain
\begin{equation}
\phi_{\mathrm{u},n}=F_n \phi_{\mathrm{c},n},
\end{equation}
with
\begin{equation}
F_n=-D^{-1}\left(K_{-1,\mathrm{u}} e^{-i k_n} +C\right).
\label{eq:f}
\end{equation}
The $\phi_{\mathrm{c},n}$ are then solutions of an effective system with reduced size
\begin{equation}
\left(K_{0}^\mathrm{eff} + K_{1}^\mathrm{eff} e^{i k_n} + K_{-1}^\mathrm{eff} e^{-i k_n}\right)\phi_{\mathrm{c},n}=0,
\label{eq:scheqe}
\end{equation}
where the effective matrices are
\begin{equation}
\begin{split}
K_{1}^\mathrm{eff}&=K_{1,\mathrm{c}} - K_{1,\mathrm{u}} D^{-1} C,\\
K_{-1}^\mathrm{eff}&=K_{-1,\mathrm{c}} - B D^{-1} K_{-1,\mathrm{u}}, \\
K_{0}^\mathrm{eff}&=A - B D^{-1} C - K_{1,\mathrm{u}}D^{-1} K_{-1,\mathrm{u}}.
\end{split}
\end{equation}
We can now solve the quadratic eigenvalue problem (eq. (\ref{eq:qep1})) for this effective system to get the set of $N_\mathrm{eff}$ eigenvectors $Q_\mathrm{c}=\left( \begin{array}{ccccc}\phi_{\mathrm{c},1}&\phi_{\mathrm{c},2}&\dots&\phi_{\mathrm{c},N_\mathrm{eff}}\end{array} \right)$ and eigenvalues $\lbrace e^{i k_n}\rbrace$ for the right-going states. The $M$ eigenvectors of the second set of solutions with $e^{i k_n}=0$ are given by $\phi_{\mathrm{c},n}=0$ with a general $\phi_{\mathrm{u},n}$. The set of eigenvectors of the full $K'$ matrix therefore is
\begin{equation}
Q=
\left( \begin{array}{cc}Q_\mathrm{c}&\mathbb{0}\\ Q_{\mathrm{u}} & Q_{\mathrm{0}}\end{array} \right),
\end{equation}
with $Q_\mathrm{u}=\left( \begin{array}{ccccc}F_1 \phi_{\mathrm{c},1}&F_2 \phi_{\mathrm{c},2}&\dots&F_{N_\mathrm{eff}}\phi_{\mathrm{c},N_\mathrm{eff}}\end{array} \right)$, and $Q_\mathrm{0}$ is a general matrix of solution vectors for the states with $e^{i k_n}=0$.
From this we obtain the set of duals
\begin{equation}
Q^{-1}=
\left( \begin{array}{cc}Q_\mathrm{c}^{-1}&\mathbb{0}\\- Q_{\mathrm{0}}^{-1}Q_{\mathrm{u}}Q_{\mathrm{c}}^{-1}& Q_{\mathrm{0}}^{-1}\end{array} \right).
\end{equation}
Using these results we can now calculate the transfer matrix $T_\mathrm{R}'$ of the transformed system
\begin{equation}
T_\mathrm{R}'=\sum_{n=1}^{N_\mathrm{eff}} e^{i k_n}\left( \begin{array}{cc} \phi_{\mathrm{c},n} \tilde{\phi}_{\mathrm{c},n}^\dagger & \mathbb{0} \\
F_n \phi_{\mathrm{c},n} \tilde{\phi}_{\mathrm{c},n}^\dagger & \mathbb{0} 
\end{array} \right),
\end{equation}
where we have also used the fact that $e^{i k_n}=0$ for the second set of solutions. We note that setting the $M$ smallest singular values $s_n$ to zero causes the last $M$ columns of $T_\mathrm{R}'$ to be zero too. Moreover the explicit calculation of $Q_\mathrm{0}$ is not needed in order to obtain $T_\mathrm{R}'$. From this and eq. (\ref{eq:sener}) we obtain the right SE
\begin{equation}
\Sigma_\mathrm{R}'=\left(\begin{array}{cc} \Sigma_{\mathrm{R}}^{\mathrm{eff}}-K_{1,\mathrm{u}}D^{-1}K_{-1,\mathrm{u}} & \mathbb{0} \\ \mathbb{0} & \mathbb{0} \end{array}\right),
\end{equation}
where
\begin{equation}
\Sigma_\mathrm{R}^\mathrm{eff}=K_1^\mathrm{eff}~\sum_{n=1}^{N^\mathrm{eff}} e^{i k_n} \phi_{\mathrm{c},n} \tilde{\phi}_{\mathrm{c},n}^\dagger
\end{equation}
is the SE of the effective system. 

%*****************************************************************
\begin{figure}
\center
\includegraphics[width=6.5cm,clip=true]{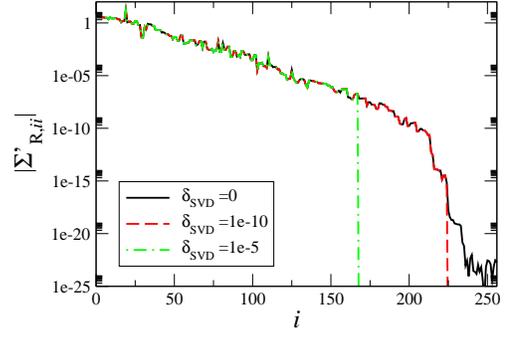}
\caption{Absolute value $|\Sigma'_{\mathrm{R},ii}|$ of the diagonal elements of the transformed right SE for different values of $\delta_{\mathrm{SVD},0}$.}
\label{fig:sigmadiag}
\end{figure}
%*****************************************************************
The structure of $\Sigma_R'$ shows that by applying this unitary transformation we have ordered the elements of the SE by absolute size, moving those columns (rows) with the smallest values to the right (bottom). By setting the smallest singular values of $K_1$ to zero those columns and rows of the SE with small values have also been set to zero. This is illustrated in figure \ref{fig:sigmadiag}, where the absolute value of the diagonal elements of the transformed selfenergy $|\Sigma'_{\mathrm{R},ii}|$ is shown for a (8,0) zigzag carbon nanotube at the Fermi energy $E_\mathrm{F}$ (see section \ref{sec:erranalysis} for a detailed description of the system). The $|\Sigma'_{\mathrm{R},ii}|$ are basically identical for different $\delta_\mathrm{SVD}$ up to $i=N_\mathrm{eff}$, and indeed an increasing value of $\delta_\mathrm{SVD}$ results in more diagonal elements of $\Sigma_R'$ set to zero. We note that $N_\mathrm{eff}$ is of similar size as $N$ in figure \ref{fig:sigmadiag}, since the system is rather short along $z$ and a small basis set is used (i.e. $N$ is small). For large systems and rich basis sets the ratio $N_\mathrm{eff}/N$ will decrease. The physical interpretation of the zero columns and rows in the SE is that the $M$ states with $k_n\rightarrow \infty$ decay infinitely fast, so that the interaction of those states is limited to the site they are localized at. Finally the SE of the original system can be obtained by applying the inverse unitary transformation
\begin{equation}
\Sigma_\mathrm{R}=U \Sigma_\mathrm{R}' U^\dagger,
\end{equation}
and in contrast to $\Sigma_\mathrm{R}'$ the matrix $\Sigma_\mathrm{R}$ is a dense $N\times N$ matrix.

Note that in order to obtain the left SE we perform the unitary transformation $K_{zz'}'=V^\dagger K_{zz'} V$, $\phi_{\mathrm{R},n}'=V \phi_{\mathrm{R},n}'$, and then follow an analogous procedure. In this case however instead of the right-going states the left-going ones are used.

\subsection{Limiting the smallest singular value}
\label{sec:svdlimit}

We can limit the lower bound of the singular values $s_n$ by setting $s_\mathrm{SVD}=s_\mathrm{max} \delta_\mathrm{SVD}$ in eq. (\ref{eq:ssvd}). In this case the approximated $K$ matrix is obtained by replacing $K_1$ with $K_{1,\mathrm{SVD}}$. The error introduced is now of the order of $s_{\mathrm{max}}~\delta_\mathrm{SVD}$. Ideally $s_{\mathrm{max}}~\delta_\mathrm{SVD}$ should be of the order of the machine numerical precision, so that the error is minimal. However sometimes increasing $s_{\mathrm{max}}~\delta_\mathrm{SVD}$ beyond that value improves the results, therefore $\delta_\mathrm{SVD}$ is left as a parameter to adjust depending on the material system investigated. This will be discussed extensively in the next section.

A simpler but equally effective possibility for limiting the smallest singular value of a matrix is that of adding a small random perturbation.\cite{teng1,ttao} Thus another strategy for reducing the condition number of $K_1$ is that of replacing $K_{1}$ with $K_{1,\mathrm{noise}}=K_{1}+W(w_{\mathrm{noise}})$, where $W(w_{\mathrm{noise}})$ is a matrix whose elements are random complex numbers with an average absolute value $|W_{ij}|\sim w_{\mathrm{noise}}$. In particular we choose the $|W_{ij}|$ in such a way that both $\mathrm{Re}(W_{ij})$ and $\mathrm{Im}(W_{ij})$ are random numbers in the range $[-w_{\mathrm{noise}},w_{\mathrm{noise}}]$. We find that if $w_{\mathrm{noise}}=s_{\mathrm{max}}~\delta_{\mathrm{SVD}}$ the addition of noise usually gives results as accurate as those obtained with the SVD procedure, but the calculation is faster since instead of performing a SVD we just perform a sum of the matrices.

\vspace{0.5cm}
In figure \ref{fig:diagram} we present our final extended algorithm as it has been implemented in \textit{Smeagol}. This now includes the following regularization procedure of $K_1$. First the size of $K_1$ and hence of the whole problem is reduced by using the scheme described in section \ref{sec:svdreduce}, with a tolerance parameter $\delta_\mathrm{SVD}=\delta_\mathrm{SVD,1}$. This generates an effective matrix $K_1^\mathrm{eff}$ whose condition number $\kappa(K_1^\mathrm{eff})$ is reduced by adding a small noise matrix $W(w_\mathrm{noise})$. Such a step is extremely fast and enhances considerably the numerical stability of the calculation. In most cases the SE for the effective system can then be calculated and no further regularization steps are needed. However, in some cases the calculation of the SE still fails. This, for example, happens when the solution of eq. (\ref{eq:qep1}) for the effective system fails, or else when the calculated number of left-going states erroneously differs from the number of right-going states. In these critical situations we further decrease $\kappa(K_1^\mathrm{eff})$ by limiting the smallest singular value of $K_1^\mathrm{eff}$ as described in section \ref{sec:svdlimit} with a tolerance parameter $\delta_\mathrm{SVD}=\delta_\mathrm{SVD,2}$. The code automatically adjusts $\delta_\mathrm{SVD,1}$, $\delta_\mathrm{SVD,2}$ and $w_\mathrm{noise}$ within a given range until the SE is calculated. In our test calculations for a number of different systems we found no situation where such a scheme has failed. In contrast when the standard algorithm of reference [\onlinecite{stefanosene}] is employed the number of failures was considerable. Note that our extended algorithm can also be used in conjunction with recursive methods for evaluating the SE.\cite{kudrnovsky1,kudrnovsky2,lopez,nardelli} Also in this case it will decrease the computing time for large systems due to the reduced size of the effective $K$ matrix.

\section{Error analysis}
\label{sec:erranalysis}

When recursive algorithms are used the accuracy of the SE is automatically known as it coincides with the convergence criterion. Poor convergence is found when the error can not be reduced below a given tolerance. Direct methods, as the one presented here, are in principle error free in the sense that when the solution is found, this is in principle exact. For this reason the numerical errors arising from direct schemes usually are not estimated. In this section we perform this estimate and present a detailed error analysis for three different material systems.

In order to estimate the numerical accuracy we use the recursive relations of eqs. (\ref{eq:rgfl}) and (\ref{eq:rgfr}), written as
\begin{equation}
\begin{split}
\Sigma_\mathrm{L}^{\mathrm{out}} &= - K_{-1} \left[K_0 +\Sigma_\mathrm{L}^\mathrm{in} \right]^{-1} K_{1}\\
\Sigma_\mathrm{R}^{\mathrm{out}} &= - K_1 \left[K_0 +\Sigma_\mathrm{R}^\mathrm{in} \right]^{-1} K_{-1},
\end{split}
\label{eq:sigmaout}
\end{equation}
where $\Sigma_\mathrm{\lbrace{L/R}\rbrace}^\mathrm{in}$ are calculated with our extended algorithm, and $\Sigma_\mathrm{\lbrace{L/R}\rbrace}^{\mathrm{out}}$ are obtained by evaluating the right-hand side term of the above equations. When the solution is exact then $\Sigma_\mathrm{L}^{\mathrm{out}}=\Sigma_\mathrm{L}^{\mathrm{in}}$ and $\Sigma_\mathrm{R}^{\mathrm{out}}=\Sigma_\mathrm{R}^{\mathrm{in}}$. Therefore we can define a measure of the error $\Delta_\Sigma$ as
\begin{equation}
\Delta_\Sigma = \left|\left| \Sigma_\mathrm{\lbrace L/R\rbrace}^{\mathrm{out}} - \Sigma_\mathrm{\{L/R\}}^\mathrm{in}\right|\right|_{\mathrm{max}},
\label{eq:error}
\end{equation}
where $\left|\left|\hdots\right|\right|_{\mathrm{max}}$ stands for the max norm,\cite{higham} the corresponding relative error is $\Delta_{\Sigma,\mathrm{r}} = \Delta_\Sigma /\left|\left|  \Sigma_\mathrm{\{L/R\}}\right|\right|_{\mathrm{max}}$.
The accuracy criterion used in the extended algorithm is the following. We first set $\delta_\mathrm{SVD,1}$, $w_\mathrm{noise}$ and eventually $\delta_\mathrm{SVD,2}$ and compute $\Delta_{\Sigma,\mathrm{r}}$. This should be lower than a target accuracy $\Delta_{\Sigma,\mathrm{r}}^\mathrm{max}$. If this is not the case then the SE will be recalculated with a different set of tolerance parameters, until $\Delta_{\Sigma,\mathrm{r}}$ reaches the desired accuracy. If this condition is never achieved the final SE is the one with to the smallest $\Delta_{\Sigma,\mathrm{r}}$ .

We now calculate the SE for different variations of the method, chosen in order to highlight the problems arising from $K_1$ and $K_{-1}$ and to show the difference between the basic method of reference [\onlinecite{stefanosene}] and the extensions presented here. There are two main differences between the two methods. The first is that here we solve eq. (\ref{eq:qep1}) without inverting $K_1$, whereas in reference [\onlinecite{stefanosene}] $K_1^{-1}$ is used to solve the inverse band-structure relation $k=k(E)$. Clearly this second choice is less accurate if $K_1$ is close to singular. However it is much faster computationally, so that it might be of advantage for big systems. The second difference is that here it is not necessary to calculate $g_{00}$ via eq.~(\ref{eq:g00a}), so that one does not need to invert $T_\mathrm{R}$ and $\bar{T}_\mathrm{R}$. 

In order to investigate the effect of these two aspects independently, we have calculated the SE using the following four methods. In method 1 we use the algorithm presented in this work. In particular we use eq. (\ref{eq:qep1}) to solve the quadratic eigenvalue problem and eqs. (\ref{eq:senel}-\ref{eq:sener}) to obtain the SE (for the right SE we actually use a different form of eq. (\ref{eq:qep1}), see Appendix C). Method 2 is essentially the same, with the only difference that instead of solving eq. (\ref{eq:qep1}) we use the eigenvalue method of reference [\onlinecite{stefanosene}]. In method 3 we solve eq. (\ref{eq:qep1}), but we use eq. (\ref{eq:sgfstefano}) to calculate the SGF, with $g_{00}$ obtained from eq. (\ref{eq:g00a}). Finally method 4 is the algorithm of reference [\onlinecite{stefanosene}].

In order to obtain a statistically significant average of the errors, we plot a histogram of the calculated errors for both $\Sigma_\mathrm{L}$ and $\Sigma_\mathrm{R}$ for a large energy range.  Here we use the absolute error, since it can readily be compared to the energy scale of the problem. Note that although the relative error might be small, the absolute error can be very large if $\left|\left|  \Sigma_\mathrm{\{L/R\}}\right|\right|_{\mathrm{max}}\gg 1$~Ry.  Furthermore in order to keep the analysis simple in all the calculations of this section we do not reduce the system size nor do we add noise ($w_\mathrm{noise}=0$). We regularize $K_1$ and $K_{-1}$ by using $s_{\mathrm{SVD}}=s_{\mathrm{max}}~\delta_\mathrm{SVD}$ in eq. (\ref{eq:ssvd}). Since the error depends on the chosen $\delta_\mathrm{SVD}$, here we calculate  $\Delta_{\Sigma}$ for a set of  $\delta_\mathrm{SVD}$ in the range $[0,10^{-23},10^{-22},\hdots,10^{-4},10^{-3}]$. We then present the smallest $\Delta_\Sigma$ found for $\delta_\mathrm{SVD}$ taken in that range. This is the smallest possible error achievable with a given method and allows us to extract informations on the range of optimal SVD values for a given method.  %*****************************************************************
\begin{figure}
\center
\includegraphics[width=7.0cm,clip=true]{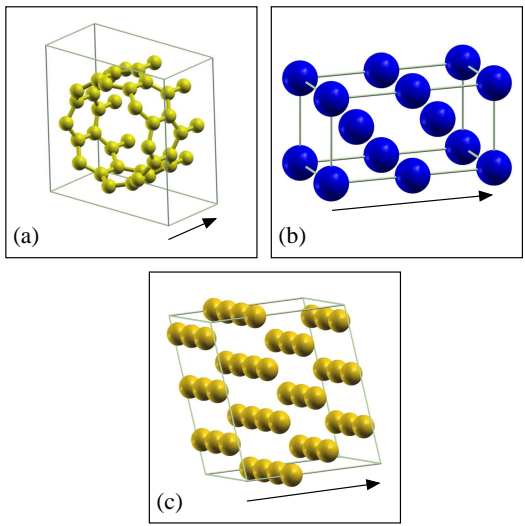}
\caption{Unit cells of the three systems investigated in this work: (a) (8,0) zigzag carbon nanotube, (b) bcc Fe oriented along the (100) direction, (c) fcc Au oriented along the (111) direction. The black arrow indicates the direction of the stacking $z$, i.e. the direction of the transport.}
\label{fig:ucellc}
\end{figure}
%*****************************************************************
%*****************************************************************
\begin{figure}
\center
\includegraphics[width=8.0cm,clip=true]{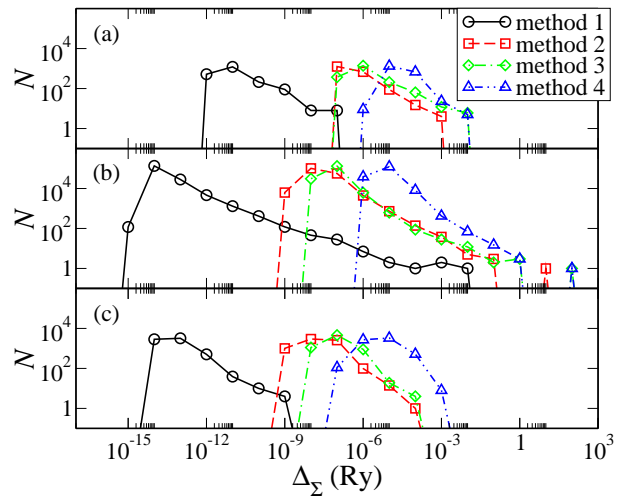}
\caption{Histogram of the errors in the calculation of the self-energy $\Delta_\Sigma$ for three different systems. (a) (8,0) zigzag carbon nanotube, (b) bcc Fe, (c) fcc Au. $N$ is the number of times a given error $\Delta_\Sigma$ occurs (not normalized).}
\label{fig:deltaall}
\end{figure}
%*****************************************************************
%*****************************************************************
\begin{figure}
\center
\includegraphics[width=8.0cm,clip=true]{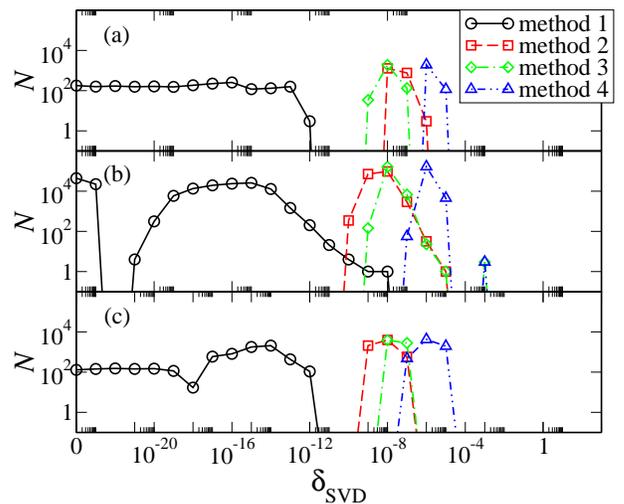}
\caption{Histogram of $\delta_\mathrm{SVD}$ giving the smallest error in the self-energy. (a) (8,0) zigzag carbon nanotube, (b) bcc Fe, (c) fcc Au. $N$ is the number of times a given $\delta_\mathrm{SVD}$ generates the smallest error (not normalized).}
\label{fig:fsall}
\end{figure}
%*****************************************************************

As first example a (8,0) zigzag carbon nanotube\cite{bulusheva} is presented (the unit cell is shown in figure \ref{fig:ucellc}(a)). The length of the periodic unit cell is 4.26 \AA~along the nanotube, with 32 carbon atoms in the unit cell.  The LDA approximation (no spin-polarization) is used for the exchange correlation potential. We consider $2s$ and $2p$ orbitals for carbon with double-$\zeta$ and a cutoff radius $r_c$ for the first $\zeta$ of $r_c=5$ Bohr. Higher $\zeta$ are constructed with the split-norm scheme with a split-norm of 15\%.\cite{siesta} The real space mesh cutoff is 200 Ry. The matrices $H_0$, $H_1$, $S_0$ and $S_1$ are extracted from a ground state DFT calculation for an infinite periodic nanotube. We calculate the SE for the semi-infinite nanotube at 1024 energy points in a range of $\pm 5$ eV around the Fermi energy.

Figure \ref{fig:deltaall}(a) shows the histogram of the errors in the SE, where $N$ is the number of times a given error $\Delta_\Sigma$ appears. In general the figure shows that for this system the average error increases when going from method 1 to method 2 and method 3, and finally to method 4. The error obtained with method 1 is on average about 6 orders of magnitude smaller than the one obtained with method 4. The main reason behind this dramatically improved accuracy is that method 1 does not involve any steps where a singular $K_1$ leads to divergencies. Method 4 on the other hand is strongly dependent on the condition number of $K_1$, since it necessitates to invert $K_1$ and $T_\mathrm{R}$ (or $\bar{T}_\mathrm{R}$). Methods 2 and 3 are on average about one order of magnitude more precise than method 4. Since they both still involve one of the two inversions the difference is however not large.

Figure \ref{fig:fsall}(a) shows the histogram of the optimum $\delta_{\mathrm{SVD}}$ used for the calculations of the SE. Here we plot the number of times $N$ a particular $\delta_\mathrm{SVD}$ has given the smallest error in the set of calculations. A larger optimal value for $\delta_{\mathrm{SVD}}$ indicates a stronger dependence of the computational scheme on $\kappa(K_1)$. For method 1 the range of used $\delta_{\mathrm{SVD}}$ is smaller than $10^{-12}$. If we force $\delta_{\mathrm{SVD}}$ to be zero we get almost the same level of accuracy as shown in figure \ref{fig:deltaall}(a), which confirms that the accuracy of for method 1 depends little on $\kappa(K_1)$ for this system. However also for this method there is a set of energies (a few percent of the total number) where the solution of eq. (\ref{eq:qep1}) fails if $\delta_{\mathrm{SVD}}$ is too small. The optimal $\delta_{\mathrm{SVD}}$ for the other methods is orders of magnitude larger than that of method 1, and it is never smaller than $10^{-9}$. The absolute error induced by replacing $K_1$ by $K_{1,\mathrm{SVD}}$ is of the order of $\delta_\mathrm{SVD}~s_{\mathrm{max}}$. Usually $s_{\mathrm{max}}$ is of the order of 1 Ry, so that the error is of the order of $\delta_\mathrm{SVD}$ Ry. Therefore since in methods 2 to 4 a large value of $\delta_{\mathrm{SVD}}$ is needed in order to improve $\kappa(K_{1,\mathrm{SVD}})$, also the resulting error is large. 

The second example is bcc Fe (figure \ref{fig:ucellc}(b)), oriented along the (100) direction. The lattice parameters are the same as in reference [\onlinecite{butler1}]. There are 4 Fe atoms in the unit cell. We apply periodic boundary conditions in the direction perpendicular to the stacking, so that these correspond to 4 Fe planes. The length of the cell along the stacking direction is 5.732 \AA. A double-$\zeta$ $s$ ($r_c$=5.6 Bohr), single-$\zeta$ $p$ ($r_c$=5.6 Bohr) and single-$\zeta$ $d$ ($r_c$=5.2 Bohr) basis is used.  The real space mesh cutoff is 600 Ry, and the DFT calculation is converged for 7x7 $k$-points in the Brillouin zone orthogonal to the stacking. The SE have been calculated for the converged DFT calculation at 32 different energies in a range of $\pm$1~eV around the Fermi energy, and for 10,000 $k$-points in the 2D Brillouin zone perpendicular to the stacking direction. For each $k$-point there is a different set of matrices $K_0$, $K_1$ and $K_{-1}$, so that for each $k$-point there is a different SE. The histogram for the error of the calculated self-energy $\Delta_\Sigma$ is shown in figure \ref{fig:deltaall}(b), and the histogram for the optimal $\delta_{\mathrm{SVD}}$ in figure \ref{fig:fsall}(b). The general behavior is similar to the one found for the carbon nanotube. We note that, although for the vast majority of the calculations the error in the SE is small, there is a long tail in the histograms of figure \ref{fig:deltaall}(b) indicating the presence of a small number of large errors. This is present for all the methods, with a maximum error of ~$10^{-2}$ Ry for method 1, and ~$100$ Ry for method 4. Closer inspection shows that the reason for the increase of the error for certain energies and $k$-points is caused by a divergence in $\left|\left|\Sigma_{\lbrace\mathrm{L,R}\rbrace}\right|\right|_{\mathrm{max}}$. This will be illustrated in more detail in the next section.

Finally we consider fcc Au (figure \ref{fig:ucellc}(c)), with the stacking along the (111) direction. The unit cell consists of three planes of nine gold atoms each. These are the typical leads used for the calculations of the transmission properties of molecules attached to gold.\cite{ratner1,ratner2,Cormac1,Cormac2} We use double-$\zeta$ $s$ ($r_c$=6.0 Bohr) and single-$\zeta$ $d$ ($r_c$=5.5 Bohr) and four $k$-points in the Brillouin zone perpendicular to the stacking. The mesh cutoff is 400 Ry. The SE have been calculated for 418 energy points, from about 15 eV below to about 10 eV above the Fermi energy. The general behavior (figures \ref{fig:deltaall}(c) and \ref{fig:fsall}(c)) is again similar to that of the previous examples. Also here the error for method 1 is about 6 orders of magnitude smaller than that of method 4, with method 2 and 3 giving some marginal improvement.

Our results show that the new scheme in general allows the calculation of the SE with high accuracy. The main advantage of method 1 is rooted in the possibility of using a much smaller $\delta_{\mathrm{SVD}}$. For big systems sometimes one might prefer to use method 2, since it is considerably faster than method 1 and gives the second best accuracy. In this case we first calculate the SE with method 2 and check the error. Only for those energy points where the error is above some maximum value (of the order of $10^{-5}$ Ry for example) the calculation is repeated with method 1 to improve the accuracy. Finally the results show that for all methods the SVD transformation of $K_1$ is necessary, although for method 1 it is needed only a few percent of the times. For big systems, in particular if the unit cell is elongated along the stacking direction, or if a rich basis set is used, $\kappa(K_1)$ will generally increase as there will be some singular values of $K_1$ going to zero. In these cases also method 1 will require a SVD transformation for most energies. The range of $\delta_{\mathrm{SVD}}$ should however be similar to the one shown in figure \ref{fig:fsall}, so that also the error in the SE should be of the same order of magnitude. We also note that in order to keep the analysis simpler here we have not used the reduction of system size described in section \ref{sec:svdreduce}, for such large systems it is however crucial in order to decrease the computational effort and regularize $K_1$ at the same time.

\section{Surface states}
\label{sec:sstates}

\begin{figure}
\center
\includegraphics[width=7.0cm,clip=true]{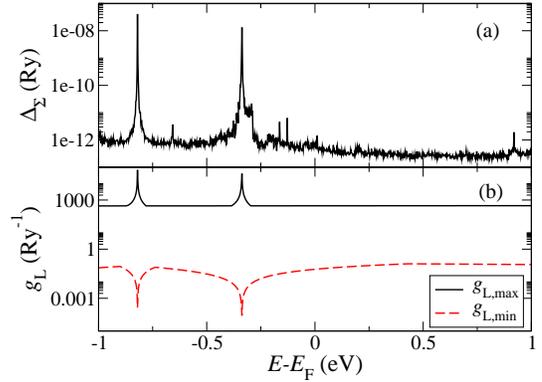}
\caption{\label{fig:ev_gf}Error analysis for the carbon nanotube of figure \ref{fig:ucellc}: (a) absolute error 
$\Delta_\Sigma$ of the self-energy as function of the energy $E$, (b) maximum ($g_{\mathrm{L},\mathrm{max}}$) 
and minimum ($g_{\mathrm{L},\mathrm{min}}$) eigenvalues of $G_\mathrm{L}$.}
\label{fig:evg_ds}
\end{figure}

The center of the error distribution for method 1 (figure \ref{fig:deltaall}) is located at small $\Delta_\mathrm{\Sigma}$, usually smaller than 10$^{-11}$ Ry. However the histogram has also a tail reaching up to very large errors. These are found only at some critical energies as demonstrated in figure \ref{fig:ev_gf}(a), where we show $\Delta_\Sigma$ for the carbon nanotube calculated over 1024 energy points in a range of 2 eV around the Fermi energy. The average error is of the order of 10$^{-12}$ Ry, but at energies around -0.8~eV and -0.34~eV the error drastically increases. Indeed a finer energy mesh at these points suggests a divergence. The origin of the large errors at particular energies can be investigated by looking at the eigenvalues $g_{\mathrm{L},i}$ of the SGF $G_{\mathrm{L}}$. In figure \ref{fig:ev_gf}(b) the largest and the smallest absolute value for the eigenvalues, respectively $g_{\mathrm{L},\mathrm{max}}$ and $g_{\mathrm{L},\mathrm{min}}$, are plotted as function of energy ($g_{\mathrm{L},\mathrm{min}} \le |g_{\mathrm{L},i}| \le g_{\mathrm{L},\mathrm{max}}$). It can be seen that $g_{\mathrm{L},\mathrm{max}}$ diverges close to the energies where the error increases, i.e. we can associate large errors in $G_\mathrm{L}$ with a divergence in its spectrum. Since $\Sigma_{\mathrm{L}}$ is calculated from eq. (\ref{eq:senel}) the only possible origin for the divergence is in the norm of some of the $\tilde{\phi}_{\mathrm{R},n}$. As these are obtained by inverting the matrix
$Q=\left(\begin{array}{cccc}
\phi_{\mathrm{R},1} &
\phi_{\mathrm{R},2} &
\hdots&
\phi_{\mathrm{R},N}
\end{array}\right)$
(eq. (\ref{eq:Q})), one deduces that the set of vectors $\lbrace\phi_{\mathrm{R},n}\rbrace$ is not linearly independent. For these energies  $\kappa(Q)\rightarrow \infty$. We therefore can simply check the magnitude of $\kappa(Q)$ to determine whether there is a divergence of the SE close to a particular energy.

Physically the divergence of the SE translates into the presence of a surface state at that particular energy.\cite{lopez,kudrnovsky1} Consider the spectral representation of $G_\mathrm{L}$
\begin{equation}
G_{\mathrm{L}}(E)=\sum_{n=1}^N \frac{1}{E+i\delta-E_n} \psi_n \tilde{\psi}_n^\dagger,
\label{eq:gldiagonal}
\end{equation}
where $E_n$ are the eigenvalues and $\psi_n$ are the right eigenvectors of the effective surface Hamiltonian matrix $H_0-\Sigma_\mathrm{L}$ with overlap $S_0$, and $\tilde{\psi}_n$ are the left eigenvectors of the same Hamiltonian. A localized surface state is found when there is a real eigenvalue $E_n(E)$ at $E_n(E)=E$ (or more generally if Im($E_n(E)$) is very small).

From the recursive relation (\ref{eq:rgfl}) one can deduce that for an infinite eigenvalue there is also a corresponding vanishing eigenvalue. Therefore in figure \ref{fig:ev_gf}(b) for energies where  $g_{\mathrm{L},\mathrm{max}}\rightarrow \infty$ we have also $g_{\mathrm{L},\mathrm{min}}\rightarrow 0$. Close to the singularity we can therefore expand the two eigenvalues as $g_{\mathrm{L},\mathrm{max}}\propto\frac{1}{E+i\delta-E_n}$ and $g_{\mathrm{L},\mathrm{min}}\propto E+i\delta-E_n$. For $E=E_n$ the largest eigenvalue in eq. (\ref{eq:gldiagonal}) is then equal to $\delta^{-1}$, and the smallest is equal to $\delta$. To avoid divergence therefore the magnitude of the $G_\mathrm{L}$ eigenvalues can be bounded to a finite value $\delta^{-1}$ by introducing a small imaginary part to the energy for energies in the vicinity of a surface state.

Another possibility for limiting the size of $g_{\mathrm{L},\mathrm{max}}$ is to bound the singular values of $Q$ from below in the same way as it is done for $K_1$ (section \ref{sec:svdlimit}). This essentially imposes the $\phi_{\mathrm{R},n}$ to be linearly independent from each other. However, with this scheme it is not possible to conserve the Green's function causality, so that the SGF might have eigenvalues lying on the positive imaginary axis. Moreover we loose control over the accuracy of the computed SGF and SE. Both these problems are avoided when using a finite $\delta$.
%*****************************************************************
%*****************************************************************
%*****************************************************************
\begin{figure}
\center
\includegraphics[width=7.0cm,clip=true]{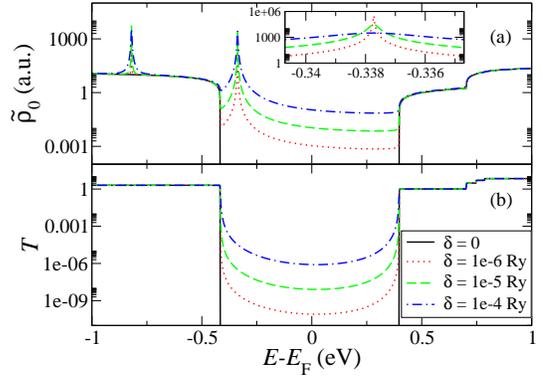}
\caption{\label{fig:trho0}Density of states and transmission coefficient for the carbon nanotube of figure \ref{fig:ucellc}.
(a) Density of states of the surface layer $\tilde{\rho}_0$ as function of energy $E$, calculated for
different broadenings $\delta$. The inset is a zoom at energies around -0.34~eV. (b) Transmission coefficient $T$ for different values of $\delta$.}
\end{figure}
%*****************************************************************

We now investigate the DOS and transport properties of a system when the finite imaginary part $\delta$ (broadening) is added to the energy. We consider as an example the carbon nanotube of figure \ref{fig:ucellc}. In figure \ref{fig:trho0}(a) the onsite surface DOS $\tilde{\rho}_0$ as defined in eq. (\ref{eq:onsitedos}) is shown for $\delta=0$~Ry, $\delta=10^{-6}$~Ry, $\delta=10^{-5}$~Ry and $\delta=10^{-4}$~Ry. For $\delta=0$ the surface DOS vanishes for energies between -0.37~eV to +0.45~eV, indicating the presence of a gap around the Fermi energy. Note that there are no Van Hove singularities in $\tilde{\rho}_0$, since we never divide by the group velocity when calculating the SGF. For finite $\delta$ and energies away from the band gap, the DOS is essentially identical to that calculated for $\delta=0$, however inside the gap $\tilde{\rho}_0$ does not vanish but saturates to a small value proportional to $\delta^{-1}$. Moreover whereas the surface states are not visible for $\delta=0$, they appear in the DOS for finite $\delta$, with a full width at half maximum (FWHM) equal to $2 \delta$.

We then move to the transport by calculating the transmission coefficient\cite{smeagol1} $T(E)$ for a carbon nanotube attached to semi-infinite leads made from an identical carbon nanotube. Since this is a periodic system $T(E)$ must equal the number of open channels, so that it can only have integer values. This is indeed the case for $\delta=0$ (figure \ref{fig:trho0}(b)). For finite $\delta$s the transmission coefficient is only approximately an integer, especially inside the energy gap region where the finite surface DOS introduced by $\delta$ leads to a non zero transmission. The transmission in the gap is proportional to $\delta^{2}$ (note that the scale is logarithmic), since on both sides of the scattering region the artificial surface DOS is proportional to $\delta$. In this region of small transmission therefore the results might change by orders of magnitude depending on the value of $\delta$. For all values of $\delta$ however we find no contributions to the transmission coming from the surface state, indicating that these do not carry current. These results show that adding a finite value $\delta$ to the energy has little effect on the actual transmission if this is large. However when the transmission is small, as in the case of tunnel junctions, the finite $\delta$ introduces an additional contribution to the conduction that might arbitrarily affect the results. It is thus imperative for those systems to identify surface states and use the imaginary $\delta$ only in a narrow energy interval around them.

Finally we can give an estimate of the relative accuracy $\Delta_{\Sigma,\mathrm{r}}(\delta)=\Delta_\Sigma/\left|\left|\Sigma\right|\right|_{\mathrm{max}}$ at the energy corresponding to the surface state. As discussed before the origin of the error is the inversion of $Q$ needed to calculate the duals. The relative error introduced by the inversion of $Q$ is proportional to $\kappa(Q)$.\cite{bautrefethen,croz,teng1,ttao,higham} Close to a surface state the smallest singular value is of the order of $\delta$, so that $\kappa(Q)\propto\delta^{-1}$. As this is the dominant source of error in the calculation of the SE close to a surface state, we can approximate the relative error as
\begin{equation}
\Delta_{\Sigma,\mathrm{r}}^\mathrm{in}=c_1~\delta^{-1},
\label{eq:dsin}
\end{equation}
where $c_1$ is a constant that depends on the machine precision and on the details of the algorithm. The label ``in'' explicitly indicates that this is the error in the SE calculated with the extended algorithm ($\Sigma_{\lbrace\mathrm{L,R}\rbrace}^\mathrm{in}$ in eq. (\ref{eq:sigmaout})). The absolute error $\Delta_\Sigma^\mathrm{in}$ is equal to the relative error times $\left|\left|\Sigma\right|\right|_{\mathrm{max}}$, which is itself proportional to $\delta^{-1}$, so that we get $\Delta_\Sigma^\mathrm{in}\propto\delta^{-2}$.

When using eq.~(\ref{eq:error}) to estimate the error in the SE we introduce an additional error due to the inversion involved in obtaining $G_\mathrm{L}$. The largest singular value of $G_\mathrm{L}$ is proportional to $\delta^{-1}$, and the smallest one is proportional to $\delta$, so that the relative error introduced by the inversion is proportional to $\kappa(G_\mathrm{L}^{-1})=\kappa(G_\mathrm{L})\propto\delta^{-2}$. For small $\delta$ we can therefore write for the error in $\Sigma^\mathrm{out}_\mathrm{L}$
\begin{equation}
\Delta_{\Sigma,\mathrm{r}}^\mathrm{out}=c_2~\delta^{-2},
\label{eq:dsout}
\end{equation}
where $c_2$ is again a constant. Since the errors are random the total estimated error can be approximated by adding the contributions from the two inversions
\begin{equation}
\Delta_{\Sigma,\mathrm{r}}^2\approx(\Delta_{\Sigma,\mathrm{r}}^{\mathrm{in}})^2+(\Delta_{\Sigma,\mathrm{r}}^{\mathrm{out}})^2.
\label{eq:errestimate}
\end{equation}
$\Delta_{\Sigma,\mathrm{r}}$ is therefore a good estimate for the true error $\Delta_{\Sigma,\mathrm{r}}^{\mathrm{in}}$ if $\Delta_{\Sigma,\mathrm{r}}^{\mathrm{out}}$ is small. Close to surface states however $\Delta_{\Sigma,\mathrm{r}}^{\mathrm{out}}\gg\Delta_{\Sigma,\mathrm{r}}^{\mathrm{in}}$, so that $\Delta_{\Sigma,\mathrm{r}}$ largely overestimates the true error.

%*****************************************************************
\begin{figure}
\center
\includegraphics[width=7.0cm,clip=true]{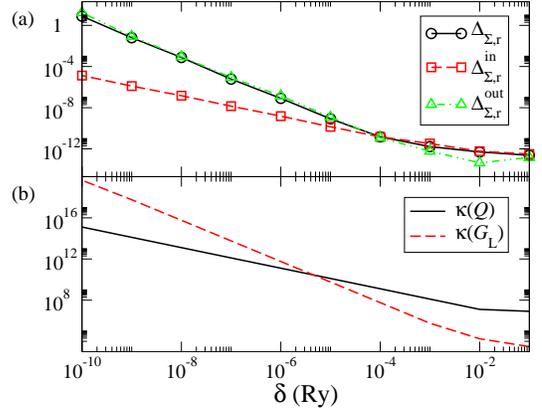}
\caption{\label{fig:ds_zoom}(a) Relative error of the self-energy ($\Delta_{\Sigma,\mathrm{r}}^\mathrm{in}$ represents the true error), (b) condition numbers of $Q$ and $G_\mathrm{L}$, as a function of the broadening $\delta$ for the carbon nanotube of figure \ref{fig:ucellc} calculated at the surface state energy.}
\label{fig:errss}
\end{figure}
%*****************************************************************

To verify these estimates numerically we present a scheme for calculating $\Delta_{\Sigma,\mathrm{r}}^{\mathrm{in}}$ and $\Delta_{\Sigma,\mathrm{r}}^{\mathrm{out}}$ independently. For each SE we perform a second calculation where we add a small amount of noise to the input matrices $K_0, K_1$, and $K_{-1}$, so that we obtain the self-energy $\Sigma_\mathrm{L,noise}$ for a slightly perturbed system. The noise is added as a random relative perturbation of each element of the matrices. As we decrease the magnitude of the noise the difference between $\Sigma_\mathrm{L}$ and $\Sigma_\mathrm{L,noise}$ is reduced until it becomes constant for noise smaller than a critical value. In this range of minimum noise even if the difference in the input matrices decreases, the difference in the output matrices is constant, it therefore corresponds to the error in the calculation. As one might expect we find that this critical value of noise is of the same order of magnitude as the numerical accuracy used (approximately $10^{-15}$ in our calculations). We can therefore obtain $\Delta_{\Sigma,\mathrm{r}}^\mathrm{in}=\left|\left|\Sigma_\mathrm{L}^\mathrm{in}-\Sigma_\mathrm{L,noise}^\mathrm{in}\right|\right|_\mathrm{max}/\left|\left|\Sigma_\mathrm{L}^\mathrm{in}\right|\right|_\mathrm{max}$ and $\Delta_{\Sigma,\mathrm{r}}^\mathrm{out}=\left|\left|\Sigma^\mathrm{out}_\mathrm{L}-\Sigma_\mathrm{L,noise}^\mathrm{out}\right|\right|_\mathrm{max}/\left|\left|\Sigma_\mathrm{L}^\mathrm{out}\right|\right|_\mathrm{max}$, with the magnitude of the noise equal to the critical value.

We have calculated the maximum error for a set of 128 energy points located within $10^{-11}$~Ry around the energy of the surface state at -0.34~eV for different values of $\delta$. The result is shown in figure \ref{fig:errss}(a). Indeed for small $\delta$ $\Delta_{\Sigma,\mathrm{r}}^\mathrm{in}$ follows eq. (\ref{eq:dsin}) with $c_1\approx10^{-15}$~Ry, $\Delta_{\Sigma,\mathrm{r}}^\mathrm{out}$ follows eq. (\ref{eq:dsout}) with $c_2\approx10^{-19}$~Ry$^2$, and $(\Delta_{\Sigma,\mathrm{r}})^2\approx(\Delta_{\Sigma,\mathrm{r}}^{\mathrm{in}})^2+(\Delta_{\Sigma,\mathrm{r}}^{\mathrm{out}})^2$. In figure \ref{fig:errss}(b) the condition numbers $\kappa(Q)$ and $\kappa(G_\mathrm{L})$ are shown, confirming $\kappa(Q)\propto\delta^{-1}$ and $\kappa(G_\mathrm{L})\propto\delta^{-2}$. This demonstrates that close to surface states $\Delta_{\Sigma,\mathrm{r}}$ is mainly caused by the calculation of $G_\mathrm{L}$. Thus $\Delta_{\Sigma,\mathrm{r}}$ largely overestimates the real error $\Delta_{\Sigma,\mathrm{r}}^\mathrm{in}$, which even for $\delta=10^{-10}$~Ry has an acceptable size of $\Delta_{\Sigma,\mathrm{r}}^\mathrm{in}\approx10^{-5}$.

Since $c_1$ and $c_2$ are generally system dependent, in practical calculations we use a value of $\delta$ ranging between $10^{-7}$ Ry and $10^{-6}$ Ry for energies in the vicinity of surface states, mainly in order to limit the absolute error. Moreover $\delta$ is added in an energy range corresponding approximately to the FWHM of the imaginary part of $(E-E_n+i \delta)^{-1}$, which is equal to $2 \delta$. Although this range is only of the order of $10^{-7}-10^{-6}$ Ry, in practical calculations where both energy and $k$-point sampling is fine the number of times when this prescription is applied can be rather large (see figure \ref{fig:deltaall}).

The above analysis confirms that close to surface states also direct methods have the same accuracy problems of recursive methods. This fact is usually ignored in the literature,\cite{stefanosene,guotaylor,umerski,butler1} where it is assumed that the accuracy is constant for a given algorithm. Here we show that the accuracy of a method is solely determined by the value of $c_1$, which, as indicated in section \ref{sec:erranalysis}, can vary over many orders of magnitude. Our analysis also shows that methods requiring the explicit calculation of $G_\mathrm{L}$ from its inverse are much less accurate close to surface states than those calculating $\Sigma_\mathrm{L}$ directly.

\section{Conclusions} \label{conclusions}

By extending the scheme proposed in reference [\onlinecite{stefanosene}] we have presented a different but equivalent form for calculating the Green's functions of an infinite quasi-1D system, as well as the SGF and SE for the semi-infinite system. We have then constructed an extended algorithm containing also the necessary steps to regularize the ill conditioned hopping matrices. This is found to be crucial in order to obtain a numerically stable algorithm. By applying a unitary transformation based on a SVD we remove the rapidly decaying states and calculate the SE for an effective system with reduced size. We further decrease the condition number of the hopping matrices by adding a small random perturbation and by limiting the smallest singular value.

We have performed a detailed error analysis on the numerical calculation of the SE, showing that if the algorithm does not involve an inversion of the hopping matrices $K_1$ (or $K_{-1}$) high accuracy is obtained. We also find that the error is not constant as function of energy. It is shown that an increase of accuracy is needed especially close to energies where the SE and SGF diverge, which corresponds to the presence of surface states in the semi-infinite system. At these energies we improved the accuracy by adding a small imaginary part to the energy. We have shown that this procedure affects the transport properties little in the high transmission limit. However, for low transmission this adds some spurious surface density of states contributing significantly to the total transmission. The transport can therefore be strongly affected, so that the imaginary part should be added only in a small energy range around the poles and it should be as small as possible.

Our final algorithm is therefore highly numerically stable and extremely accurate. Most importantly errors and accuracy can be closely monitored. We believe that this is an ideal algorithm to be used with {\it ab initio} transport schemes, where the condition of the Hamiltonian and its sparsity is controlled by the convergence of the electronic structure and therefore cannot be fixed {\it a priori}.

\acknowledgments{This work is sponsored by Science Foundation of Ireland under the grants SFI02/IN1/I175 and 
SFI07/RFP/PHY235. Authors wish to acknowledge ICHEC and TCHPC for the provision of computational facilities and support.}

\section*{APPENDIX A: VERIFICATION OF THE RECURSIVE RELATION FOR THE SGF}
\label{appendix:pv}

Here we demonstrate that $\Sigma_\mathrm{L}$ calculated using eq.~(\ref{eq:senel}) indeed fulfills the recursive relation for $G_\mathrm{L}$ of eq.~(\ref{eq:rgfl}). Insert eqs.~(\ref{eq:gfs}) and (\ref{eq:senel}) into eq.~(\ref{eq:rgfl}) and take the inverse to obtain
\begin{equation}
K_0 + K_{-1} \bar{T}_\mathrm{R}+K_{1} \bar{T}_\mathrm{R}^{-1}=0.
\end{equation}
Using the definition of the matrix $\bar{T}_\mathrm{R}$ (eq. (\ref{eq:bara})) we write
\begin{equation}
\sum_{n=1}^N \left(K_0 + K_{-1}e^{-i \bar{k}_n} +K_{1} e^{i \bar{k}_n}\right) \bar{\phi}_{\mathrm{R},n} \tilde{\bar{\phi}}_{\mathrm{R},n}^\dagger=0.
\end{equation}
This equation corresponds to the defining equation for the $\bar{\phi}_{\mathrm{R},n}$ and is therefore fulfilled by definition.  The same is therefore true for eq.~(\ref{eq:rgfl}). Eq.~(\ref{eq:rgfr}) for $G_\mathrm{R}$ can be demonstrated similarly.

\section*{APPENDIX B: REGULARIZATION OF $K_1$ AND $K_{-1}$ FOR $K_1^\dagger\ne K_{-1}$}
\label{sec:appendixnh}

In section \ref{sec:svdreduce} we assume that $K_1=K_{-1}^\dagger$ in order to write the transformed matrices $K_1'$ and $K_{-1}'$ in form of eq.~(\ref{eq:transfmat}). If $K_1^\dagger\ne K_{-1}$ the same can be done by performing a generalized SVD of the Hamiltonian and overlap matrices as described in reference [\onlinecite{smeagol1}]. Here we present a different approach, based on two standard SVD transformations, one for $K_1$ and one for $K_{-1}^\dagger$
\begin{equation}
\begin{split}
K_1&=U_1 S_\mathrm{a} V_1^\dagger,\\
K_{-1}^\dagger&=U_{-1} S_\mathrm{b} V_{-1}^\dagger.
\label{eq:k1m1svd}
\end{split}
\end{equation}
Here $U_1, U_{-1}, V_1$ and $V_{-1}$ are unitary matrices, $S_\mathrm{a}$ and $S_\mathrm{b}$ are diagonal matrices with the singular values on the diagonal. In general there are $M_1$ singular values of $K_1$ smaller than $\delta_\mathrm{SVD} s_{\mathrm{a},\mathrm{max}}$, and  $M_{-1}$ singular values of $K_{-1}$ smaller than $\delta_\mathrm{SVD} s_{\mathrm{b},\mathrm{max}}$, with $s_{\mathrm{a},\mathrm{max}}$ and $s_{\mathrm{b},\mathrm{max}}$ being respectively the largest singular value of $K_1$ and $K_{-1}$. If $M=\mathrm{min}(M_1,M_{-1})$, we obtain $K_{1,\mathrm{SVD}}$ by setting the smallest $M$ singular values of $K_1$ to zero. In the same way we obtain $K_{-1,\mathrm{SVD}}$ by setting the smallest  $M$ singular values of $K_{-1}$ to zero.  A transformation
\begin{equation}
\begin{split}
K_1'&=U_1^\dagger K_{1,\mathrm{SVD}} U_{-1},\\
K_{-1}'&=U_{1}^\dagger K_{-1,\mathrm{SVD}} U_{-1}
\end{split}
\end{equation}
brings both $K_1'$ and $K_{-1}'$ to the form of eq. (\ref{eq:transfmat}). All the results of section \ref{sec:svdreduce} are then valid also for $K_1^\dagger\ne K_{-1}$.

If the Hamiltonian and overlap matrices are real and Hermitian, but the energy is complex, then $K_1 = K_{-1}^{\dagger *}$. By using eq.~(\ref{eq:k1m1svd}), and the fact that $S_\mathrm{a}$ and $S_\mathrm{b}$ are real, we obtain $S_\mathrm{a}=S_\mathrm{b}$, so that $M=M_1=M_{-1}$. If the Hamiltonian and overlap matrices are Hermitian but not real, then in general $S_\mathrm{a}\ne S_\mathrm{b}$. However in all the calculations performed the difference between $S_\mathrm{a}$ and $S_{\mathrm{b}}$ was very small, so that in practice we always had $M_1=M_2$.

In section \ref{sec:svdlimit} we limit the singular values of $K_1$ from below without reducing the size of the system. If $K_1^\dagger\ne K_{-1}$ we simply apply the transformations described in section \ref{sec:svdlimit} to both $K_1$ and $K_{-1}$ independently.

\section*{APPENDIX C: QUADRATIC EIGENVALUE PROBLEM FOR THE RIGHT-GOING STATES}

We find that in the solution of eq. (\ref{eq:qep1}) the numerical accuracy for those eigenvalues with $|e^{i k_n}|>1$ (Im$(k_n) < 0$) is better than for those with $|e^{i k_n}|<1$ (Im$(k_n) > 0$), especially when $|k_n| \gg 1$. For $\Sigma_\mathrm{L}$ we only need the left-going states, for which eq. (\ref{eq:qep1}) gives the better accuracy. For $\Sigma_\mathrm{R}$ the right-going states are needed. In this case, in order to increase the accuracy for the right decaying states (Im$(k_n) > 0$), instead of eq. (\ref{eq:qep1}) we solve the equivalent equation
\begin{equation}
\left( \begin{array}{cc}
-K_0 &-K_{1} \\
\mathbb{1}  &\mathbb{0}
\end{array} \right) \Phi_{\mathrm{R},n}=
e^{-i k_n}
\left( \begin{array}{cc}
K_{-1} & \mathbb{0}\\
\mathbb{0} & \mathbb{1}
\end{array} \right)\Phi_{\mathrm{R},n},
\label{eq:qepr}
\end{equation}
with
\begin{equation}
\Phi_{\mathrm{R},n}=
\left( \begin{array}{c} e^{-i \frac{k_n}{2}} \\ e^{i \frac{k_n}{2}} \end{array} \right)\frac{\phi_{\mathrm{R},n}}{\sqrt{v_n}}.
\end{equation}
The eigenvalues of the states with Im$(k_n) > 0$ now have an absolute value larger than one and therefore a higher accuracy.


%\bibliography{refs_mnas,refs_femgo,refs_sene}

\end{document}